\documentclass[11pt,twoside]{article}

\usepackage{asp2006}
\usepackage{epsf}
\usepackage{psfig}
\usepackage{lscape}
\usepackage{graphicx}
\usepackage{times}

\newcommand{\kms}{km~s{$^{-1}$}}
\newcommand{\hii}{H{\sc ii}}

\newcommand{\Hi}{H{\sc i}}
\newcommand{\Ha}{\mbox{H$\alpha$}}

\begin{document}
\title{Overview of the Orion Complex}   
\author{John Bally}   
\affil{Center for Astrophysics and Apace Astronomy, University of Colorado \\
Boulder, CO 80389, USA}    

\begin{abstract}
The Orion star formation complex is the nearest region of on-going
star formation that continues to produce both low and high mass stars.
Orion is discussed in the larger context of star formation in the
Solar vicinity over the last 100 Myr.  The Orion complex is located on
the far side of the Gould's Belt system of clouds and young stars
through which our Solar system is drifting.  A review is given of the
overall structure and properties of the Orion star forming complex,
the best studied OB association.  Over the last 12 Myr, Orion has
given birth to at least ten thousand stars contained in a half dozen
sub-groups and short-lived clusters.  The Orion OB association has
been the source of several massive, high-velocity run-away stars,
including $\mu$ Columbae and AE Auriga.  Some of Orion's most massive
members died in supernova explosions that created the 300 pc
diameter Orion / Eridanus super-bubble whose near wall may be as close
as 180 pc.  The combined effects of UV radiation, stellar winds, and
supernovae have impacted surviving molecular clouds in Orion.  The large Orion
A, IC~2118 molecular clouds and dozens of smaller clouds strewn
throughout the interior of the superbubble have cometary shapes
pointing back towards the center of the Orion OB association.  Most
are forming stars in the compressed layers facing the bubble interior.
\end{abstract}

\section{Introduction}

Orion is the best studied region of star formation in the sky.  
Its young stars and gas provide important clues about the physics of 
star formation, the formation, evolution, and destruction of star 
forming clouds, the dynamics and energetics of the interstellar medium, 
and the role that OB associations and high mass stars play in the cycling of 
gas between the various phases of the ISM.   

In this chapter, we start with an overview of the Solar vicinity that establishes
the context in which nearby star forming regions must be understood.  Then
we discuss the Orion complex of star forming regions as an example of
star formation in an OB association.

\section{Spiral Arms, Superbubbles, GMCs, and Star Formation in the Solar Vicinity} 

The Solar vicinity (within about 0.5 to 1 kpc of the Sun) is the only 
place where it is possible to investigate the distributions and 
motions of young stars and associated gas in 3 dimensions and to probe
the history of star formation and the interstellar medium (ISM).
While \hii\ regions and low mass T-Tauri stars trace the youngest 
($<$ 3 to 5  Myr old) sites of on-going or recent star formation,  
OB associations can trace star formation history back  nearly 100 Myr.  
The least massive stars that end their lives in supernova
explosions have masses of about 8 M$_{\odot}$, main-sequence lifetimes 
of about 40 Myr, and spectral type B3 which can excite small HII regions.  Thus, 
groups containing B3 and earlier type stars trace sites of star birth younger
than 40 Myr.   OB associations whose most massive members have later 
spectral types can  identify locales where stars formed more than 40 Myr ago.  
The mass spectra,  locations, velocities,  and ages of young stars and the 
properties of gas  in nearby associations provide clues about the history of 
star formation  and the  origin,  evolution, and destruction of molecular clouds 
over the last 100  Myr.   Such studies enable us to
decode the recent history of the interstellar medium and associated 
star birth in our portion of the Galaxy. 

Currently, the Solar vicinity appears to be located in an inter-arm
spur of gas and dust between two major spiral arms of the Galaxy.
Looking towards the Galactic anti-center, the Perseus Arm lies about 2
kpc beyond the Solar circle.  Many well-known star forming complexes
are embedded in this arm.  They include giant cloud complexes near
{\em l}~= ~111\deg\ that contain NGC 7538 and the Cas A supernova
remnant, the W3/W4/W5 complexes near {\em l}~=~134\deg , and the
Auriga and Gem OB1 clouds in the anti-center direction (see chapters
by Kun et al., Megeath et al., and Reipurth \& Yan).  Looking towards
the inner Galaxy, the Sagittarius Arm is located about 2 kpc inside
the Solar circle and contains the M8, M16, and M17 star forming
complexes, see chapters by Tothill et al., Oliveira, and Chini \&
Hoffmeister.  Millimeter-wavelength molecular absorption against the
Galactic center clouds shows that the Sagittarius arm has a
blueshifted radial velocity of $V_{LSR} = -15$ to $-20$ \kms .  The
more distant Scutum and 3 kpc Arms appear at $V_{LSR} = -35$ and $-60$
\kms .

Superimposed on Galactic differential rotation, most of the ISM in the
Solar vicinity (GMCs and \Hi ) is expanding with a mean velocity of
${\rm 2~to~5~km~s^{-1}}$ from a point located near $l$~=~150$\rm ^o$,
$b$~=~0$\rm ^o$, d~=~200 pc, the approximate centroid of the 50 Myr old
Cas--Tau group, a ``fossil'' OB association (Blaauw 1991).  This
systematic expansion of the local gas was first identified by Lindblad
(1967, 1973) and is sometimes called `Lindblad's ring' of \Hi , but it
is even more apparent in the kinematics of nearby molecular gas (Dame
et al. 1987, 2001; Taylor, Dickman, \& Scoville 1987; Poppel et
al. 1994).  The Sun appears to be well inside this expanding ring.
The nearest OB associations, such as Sco-Cen (d $\approx$ 150 pc), Per
OB2 (d $\approx$ 300 pc), Orion OB1 (d $\approx$ 400), and Lac OB1b (d
$\approx$ 500 pc), and the B and A stars that trace the so-called
`Gould's Belt' of nearby young and intermediate age stars are all
associated with Lindblad's ring (Lesh 1968; De Zeeuw et al. 1999).
Lindblad's ring appears to be a 30 to 60 Myr old fossil supershell
driven into the local ISM by the Cas--Tau group and the associated
$\alpha$ Persei cluster (Blaauw 1991).  The Gould's Belt of stars,
nearby OB associations, and star forming dark clouds may thus
represent secondary star formation in clouds that condensed from the
ancient Lindblad ring supershell.  Figure 1 shows a schematic face-on
view of the Solar vicinity. Support for this view comes from the
agreement between the observed radial velocity fields of the local
\Hi\ emission and nearby CO clouds (Figure 2) with models of an
expanding and tidally sheared 30 to 60 Myr old superbubble powered by
the Cas-Tau group (Poppel et al. 1994).

\begin{figure*}
\center{\includegraphics[width=1.0\textwidth]{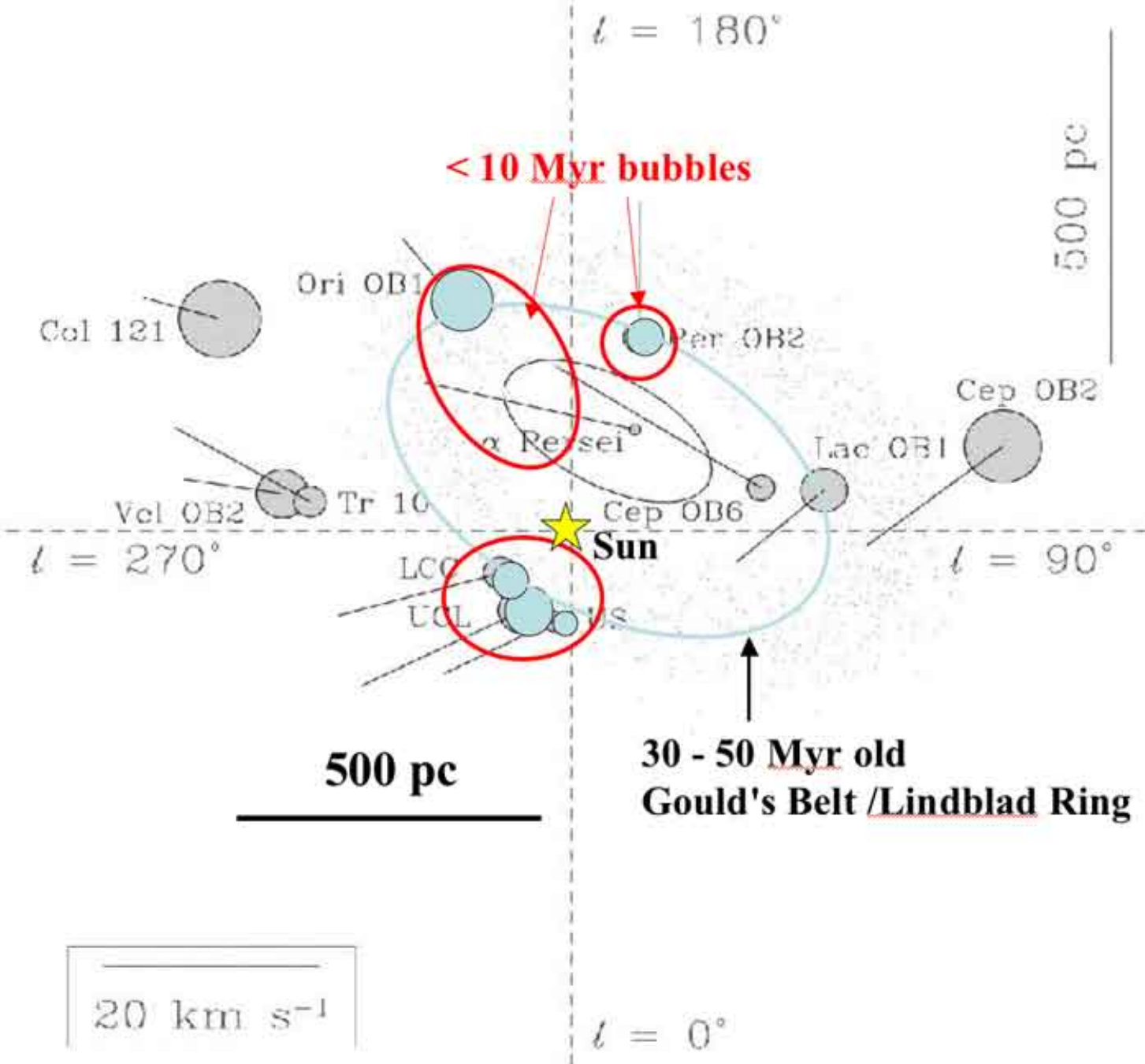}}
\caption{\small
A Sun-centric schematic face-on view of the Solar vicinity showing the older 
OB associations from
de Zeeuw et al. (1999; solid grey disks), the approximate location of the Lindblad Ring, 
(blue oval) and the  three major OB associations younger than about 20 Myr; 
Sco-Cen, Per OB2, and Orion OB1 (solid blue disks) .  The approximate outer 
boundaries of the supershells powered by each young association are shown as  
red ovals.  The Cas-Tau fossil OB association whose center is marked by the 
$\alpha$-Persei cluster, is marked by the grey oval inside the Gloul's Belt / Lindblad
Ring.
}
\end{figure*}

\begin{figure}
\center{\includegraphics[width=1.0\textwidth]{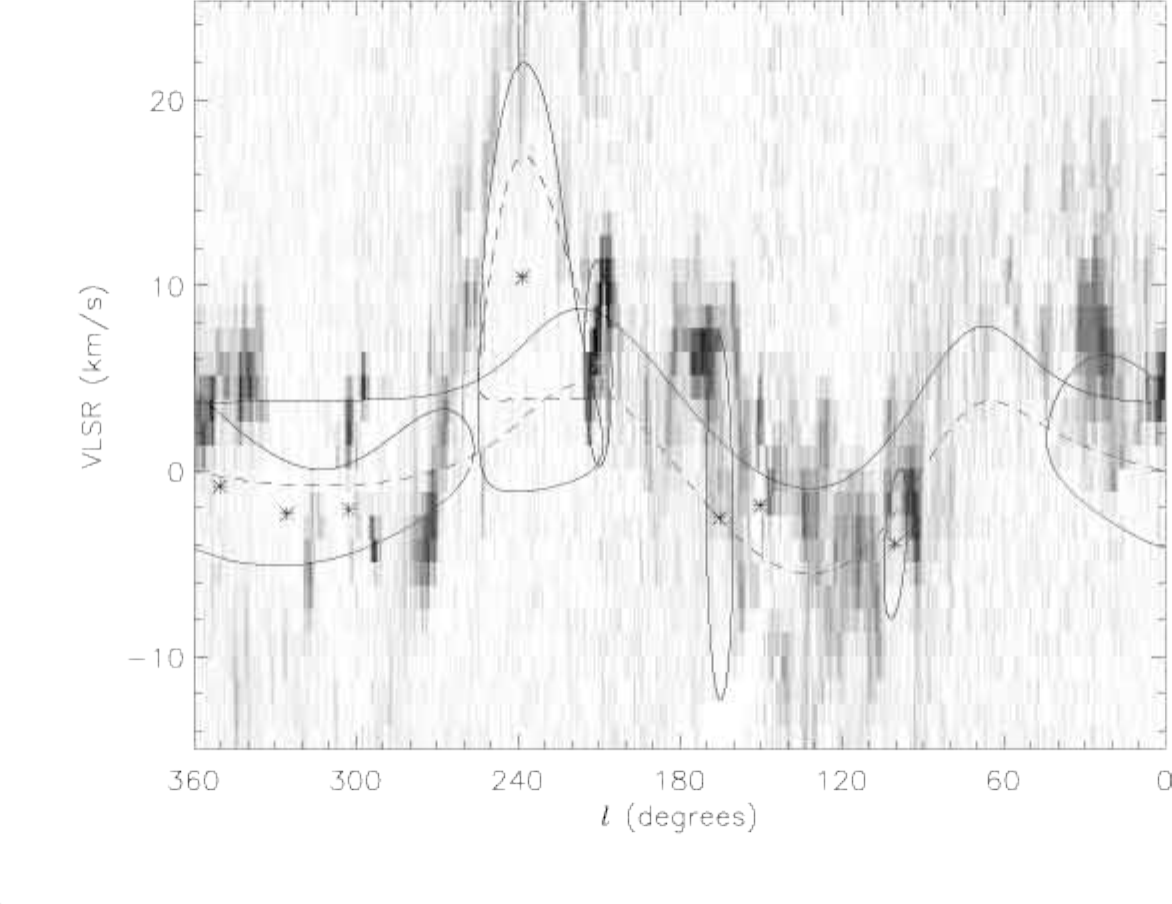}}
\caption{
The location of CO clouds in the $l$--$v$ plane with brightness
temperature shown in grey-scale (the CO data are from Dame et
al. 1987). To reduce confusion from distant gas in the disk, emission
with $\mid b \mid < 2.0\rm ^o$ is not shown. The location of the
nearby OB associations are shown by asterisks (*).  Association
velocities are derived from their longitudes, distances, and the
Galactic rotation curve.  Included are Orion at $l = 210\rm ^o$,
Sco-Cen at $l = 300\rm ^o$ to $360\rm ^o$, Per OB2 at $l = 165\rm ^o$,
Lac OB1 at $l = 100\rm ^o$, the $\alpha$ Persei cluster at
$l = 150\rm ^o$, and the expected location of the association that
created GSH238-00+9 (Heiles 1998).   Models of the supershell powered by the
``fossil'' Cas-Tau group (centered near the $\alpha$ Persei cluster)
are shown by two curves extending horizontally across the figure.
Models of the younger supershells are shown by the closed loops.
Dashed lines show the expected velocity fields of stationary 
supershells assuming that their velocities reflect {\it only} 
the motion produced  by Galactic differential rotation (e.g. the 
shells are not expanding).  The solid lines show the velocity fields 
produced by both {\it Galactic differential rotation} and {\it expansion} 
powered by the energy released by OB stars.   For the
Gould's Belt/Lindblad ring, an expansion velocity of 4.5 $\rm
km~s^{-1}$ is assumed from a point located about 170 pc from the Sun
towards $l = 131\rm ^o$. While this is 20 degrees from the estimated
center of the Cas--Tau fossil association, the original position of
Cas--Tau is uncertain due to its 200 pc extent along the Galactic
plane at a distance of 160 pc and because a small ($\approx4$ \kms)
peculiar velocity could have moved it this angular distance during the
past 15 Myr. The expansion velocities used for the other bubbles are
5.0 $\rm km~s^{-1}$ for the ``NSB'' (GSH238+00+09) identified by
Heiles at $l=240\deg$, 5.0 $\rm km~s^{-1}$ for Sco-Cen, 5.0 $\rm
km~s^{-1}$ for Orion, 10.0 $\rm km~s^{-1}$ for Perseus, and 4.0 $\rm
km~s^{-1}$ for Lac~OB1. 
}
\end{figure}

The distribution of dust in the COBE and IRAS data and the kinematics
of high-latitude \Hi\ provide additional evidence for an ancient
super-shell created by a super-bubble centered on the Cas--Tau group
that swept-up the local ISM, and blew out of the Galaxy orthogonal to
the Galactic plane.  The lines of sight with the lowest column
densities of \Hi\ (the Lockman Hole; Figure 12 in Stark et al. 1992)
and dust (Baade's Hole) lie above and below the Cas--Tau group near $l
~=~150\rm ^o$ and $b ~ \pm ~ 35\rm ^o$, providing evidence that an
ancient bubble burst out of the Galactic plane at about one dust
scale-height above and below the Cas--Tau ``fossil'' OB association.

Gas expelled from the Galactic disk 20 to 50 Myr ago is now expected
to be falling back.  Indeed, the 21~cm line profiles formed by
averaging all \Hi\ emission produced in both the northern and southern
Galactic hemispheres show excess emission at low negative velocities
in the range $\rm v_{\rm lsr}$~=~--10 to --40~\kms\ (Stark et
al. 1992) indicating an excess of infalling material at low
velocities.  Since the ambient pressure of the ISM orthogonal to the
Galactic plane declines as an exponential, an expanding pressure
driven supershell will accelerate once its radius becomes larger than
the gas layer scale-height.  As a supershell bursts out of the
Galactic plane, it is subject to Rayleigh-Taylor fragmentation
instabilities (MacLow \& McCray 1987).  After blow--out, the resulting
dense clumps will move ballistically in the gravitational potential of
the Galactic disk.  Up to about 500~pc above the plane, the
gravitational field of the disk is reasonably well represented by a
harmonic oscillator potential with a z--oscillation time of about 80
to 100 Myr (Spitzer 1978).  Clumps ejected from near the Galactic
plane at less than 40 \kms\ will stop at heights of less than 500~pc
within 20 -- 25 Myr of their formation.  During the next 20 -- 25 Myr,
they fall back towards the plane.  Thus, the dynamical age of the low
velocity infalling \Hi\ is comparable to that of the Cas--Tau group.
Figures 3 and 4 show a schematic view of super-bubble, super-shell,
and super-ring evolution.

In summary, the Sun appears to be in the interior of an ancient supershell
that may have produced the parent clouds from which Orion and most of
the other nearby star forming regions originated.  There are several lines of
evidence to support this view:

$\bullet$
The expanding network of HI clouds (the Lindblad Ring) and associated 
molecular clouds.

$\bullet$
The Gould's Belt of young stars that appears to be associated with the
Lindblad Ring.

$\bullet$
The locations of the lowest column densities of dust and HI are situated
above and below the centers of the ring.

$\bullet$
The excess of infalling,  low negative velocity HI towards the North and
South polar caps of the Galaxy.

$\bullet$
The presence of the Cas-Tau fossil OB association and its central
cluster, the $\alpha$-Per group.

The age of the Cas-Tau group, estimated to be between 40 to 90 Myr,
may indicate that it formed when the Solar vicinity made its last passage
through a major spiral arm of the Galaxy.  Since the density
of gas is higher in an arm, it is likely that so too was the star formation 
rate.    Thus,  the Cas-Tau group may have produced more stars and
a larger super-bubble than the second generation clouds and OB associations
such as the Sco-Cen and Orion OB associations, that were spawned from its 
super-ring.   The stellar content of Cas-Tau remain poorly determined.  Future
parallax, proper motion, radial velocity, and spectroscopic surveys are needed
to establish membership.

\begin{figure*}
\center{\includegraphics[width=0.9\textwidth]{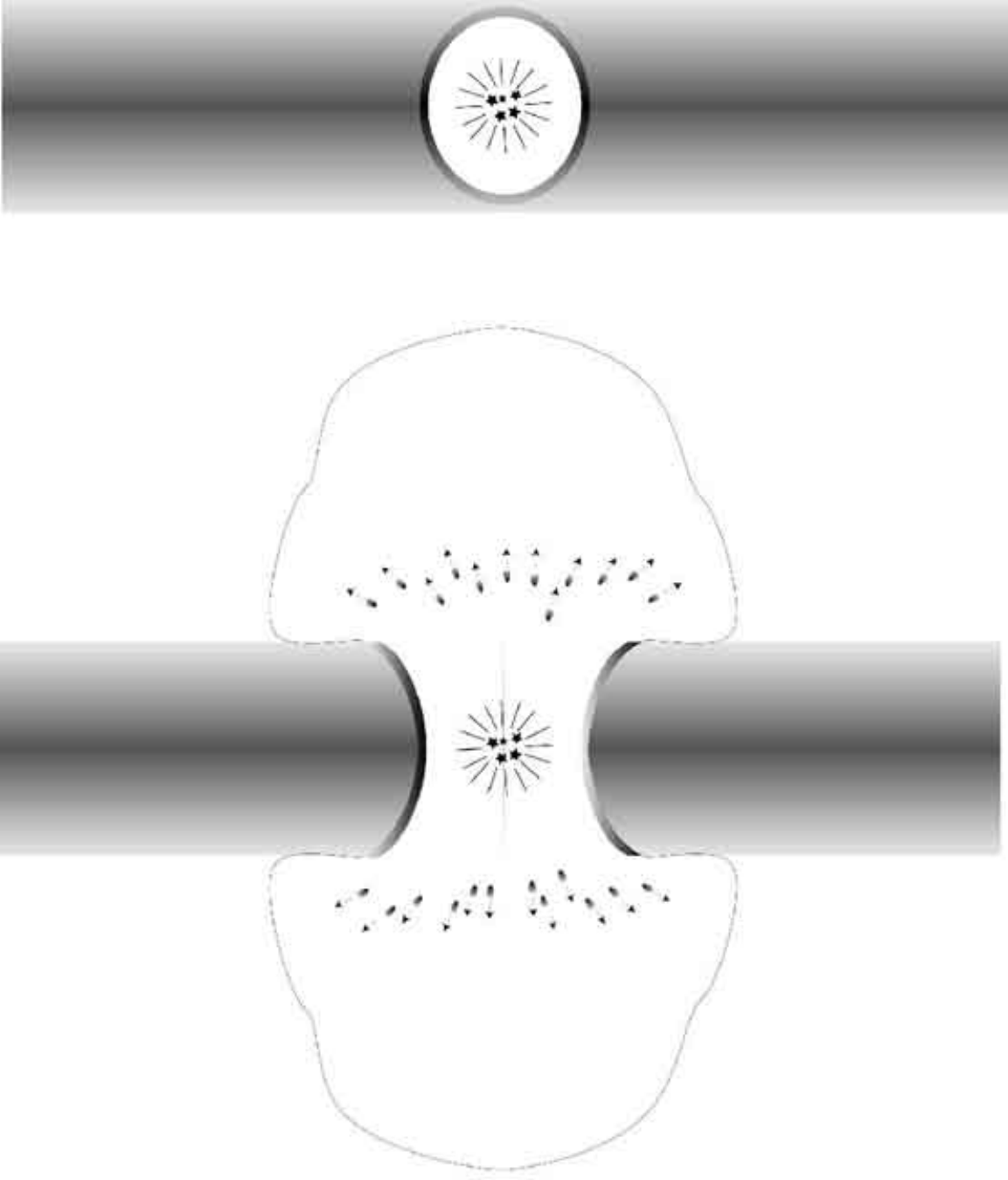}}
\caption{\small
A schematic cartoon showing a super-bubble blowing out
of the Galaxy during its early evolution (top).  UV radiation, stellar
winds, and multiple supernova explosions in the parent OB association
inject energy at a roughly steady rate for about 40 Myr as stellar
mortality depopulates the upper-end of the mass function down to a
mass of 8 M$_{\odot}$, the least massive star that can explode.  The
hot expanding bubble displaces ambient gas and sweeps it into a dense
shell.  As the swept-up shell reaches a radius larger than the
scale-height of the gas layer, it blows out of the Galactic plane
(bottom).  The densest part of the shell forms a ring in the plane of
the gas layer.  As the shell expands into the exponentially declining
density gradient above the plane, Rayleigh-Taylor instabilities can
cause the shell to fragment into dense clumps at high latitudes. }
\end{figure*}

\begin{figure*}
\center{\includegraphics[width=0.9\textwidth]{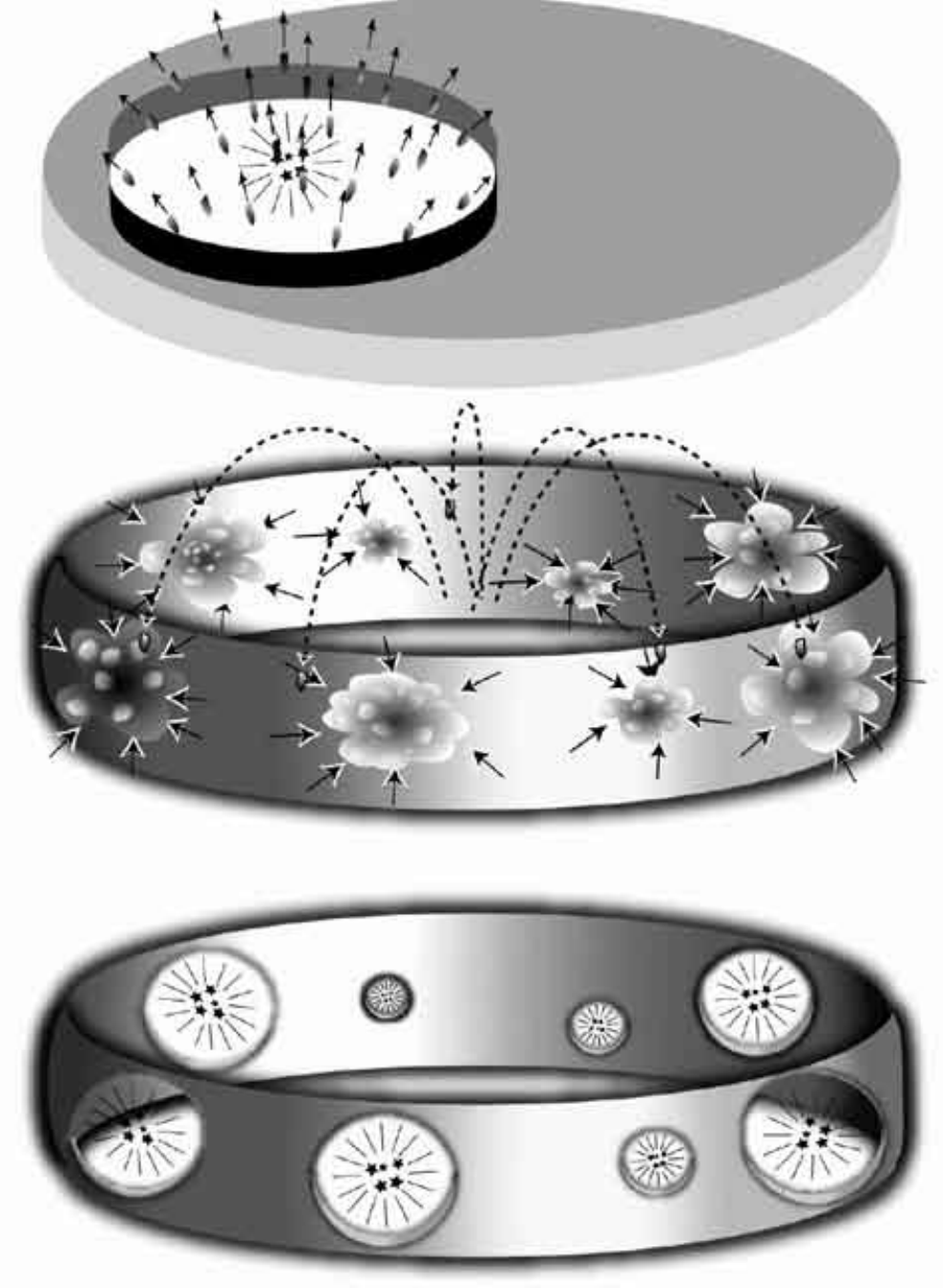}}
\caption{\small
A schematic cartoon showing the late phases in the
evolution of a super-bubble-driven super-shell and ring.  Dense clumps
formed at high latitudes (top) by instabilities in the expanding shell 
start to fall back towards the Galactic
plane after 20 to 25 Myr due to the gravitational potential of the
Galactic disk.  To first order, the gravitational field of the disk in the Solar
vicinity is well represented by a harmonic oscillator potential in the vertical 
direction that gives rise of oscillations about the mid-plane with a period 
of about 80 to 100 Myr.  Thus, clumps created by fragmentation of the
shell will tend to fall-back into the mid-plane  in about 40 to 50 Myr.  
As the shell  sweeps-up the ISM in the Galactic mid-plane and 
decelerates, the expanding ring of gas  can become unstable
to gravitational collapse (middle).  This occurs when the local
spreading velocity caused by shell expansion drops below the
gravitational escape speed from that portion of the ring.  For typical
OB association and super-bubble parameters in the Solar vicinity,
instabilities set in at a ring-age of about 30 to 50 Myr after
formation.  The first self-gravitating objects tend to have masses of
order $10^5$ M$_{\odot}$ (McCray \& Kafatos 1987).  Note the
similarity of time-scales on which high latitude debris returns to
the plane and the onset of gravitational instabilities.  These clumps 
may soon evolve into GMCs that form their own OB associations and 
new super-bubbles (bottom).
}
\end{figure*}

\section{Stars in the Orion Complex}

The Orion OB association consists of a sequence of stellar groups
of different ages that are partially superimposed along our line-of-sight  (see chapter by
Briceno and Figures 5 and 6).    Traditionally,  OB sub-group boundaries have
been drawn to segregate each into a well defined and contiguous regions on the
plane of the sky (see chapter by Brice\~no).     However, when various stellar aggregates  
are superimposed on the  plane of the sky,  it make sense to also incorporate ages into  
the sub-group classification.    Although there are differences in the estimated 
ages of the various groups, most workers agree that the oldest group, Orion OB1a,  
is  located  northwest of  Orion's Belt,  and has an age of about  8 to  12 Myr
(Blaauw 1991; Brown et al. 1994).   The  OB1b subgroup is 
centered on the Belt and has been estimated to have an age
ranging from 1.7  to 8 Myr.    However, the younger age is inconsistent with
the presence of the three supergiants that form the naked-eye Belt stars which must
be at least 5 Myrs old given their masses.   A recently discovered cluster of roughly 7 to 
10 Myr old stars is centered around 25 Ori at the northwestern end of Orion's Belt 
(Brice\~no et al. 2007).  Although formally a part of the Orion OB1a sub-group, the 
25 Ori cluster has a distinct radial velocity, being about 10 \kms\ lower than the 
traditional members of the 1a subgroup.  It has been proposed
that the 25 Ori group was formed as the HII region created by the 1a subgroup
expanded into surrounding gas and triggered a burst of star formation.  
The 25 Ori group may thus represent a group that formed between Ori 1a and 1b.

The 2 to 6 Myr old OB1c subgroup consists of stars located in  Orion's Sword 
about 4$\rm ^o$ below the Belt and  directly in front  of the Orion Nebula.   This
subgroups contains two loose clusters, NGC 1980 at the southern end of the Sword,
and NGC 1981 at the northern end (Figure~6).    The older stars in the Sword 
are superimposed on the much younger stellar populations associated with the Orion
Nebula, M43, NGC 1977, and the OMC1, 2, and 3 regions in the Integral Shaped 
FIlament at the northern end of the Orion A molecular cloud (see Figure 2 in the Chapter
by O'Dell et al.).   Thus, it is hard to separate these two stellar populations and it is 
unclear weather the 1c and 1d sub-groups represent different  populations, or merely
older and younger stellar groups that formed from the Orion A cloud at different times.
See the chapter by Muench et al. for further discussion.

The $\lambda$ Ori group (see chapter by Mathieu) may also have been triggered by
the expansion of the bubble created by Orion OB1a.   This group has an age similar to 
the Orion OB1b or the older stars in OB1c and, as illustrated in Figure 5, the 
$\lambda$ Ori group is located at approximately the  same distance from the 
center of Ori OB1a as Ori OB1c   Thus, the $\lambda$ Ori  group could be 
considered to be a disjoint  portion of OB1c.

\begin{figure*}
\center{\includegraphics[width=1.1\textwidth]{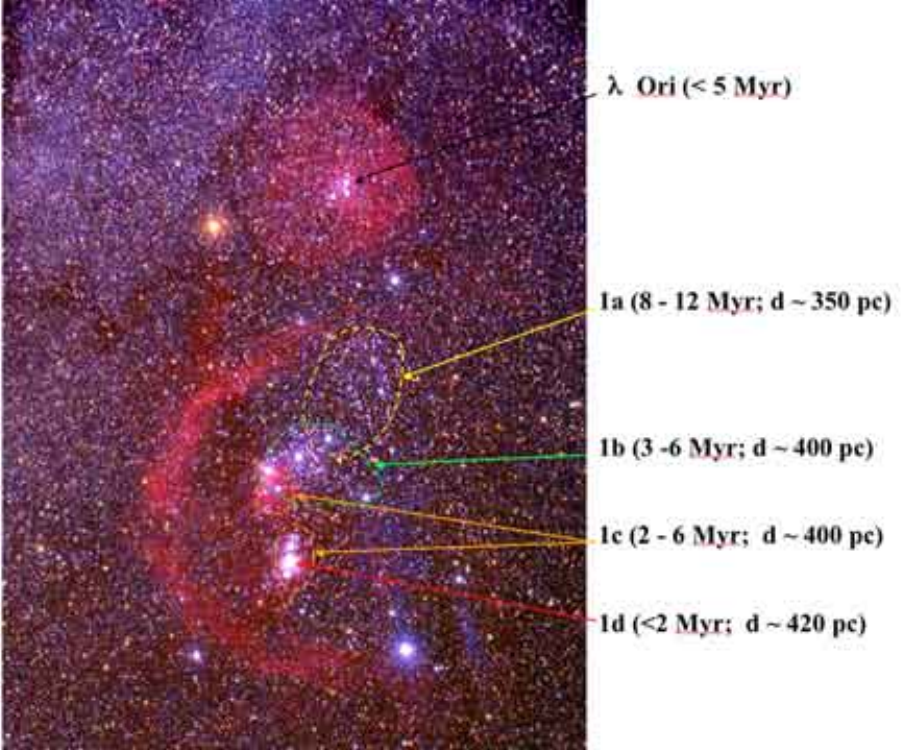}}
\caption{\small
A wide field image showing Orion with the various sub-groups of 
the Orion OB1 association.  
}
\end{figure*}

\begin{figure*}
\center{\includegraphics[width=1.1\textwidth]{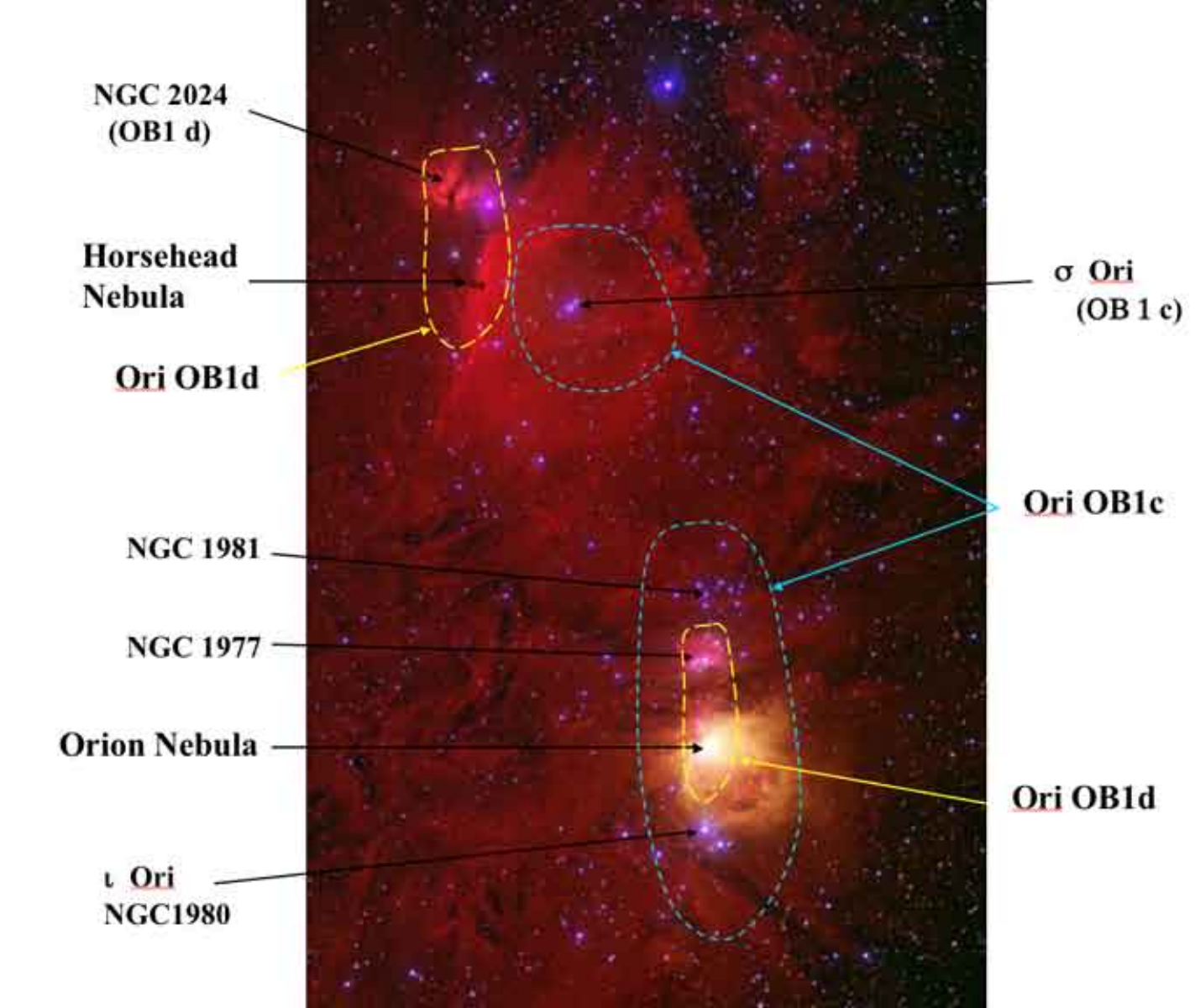}}
\caption{\small
A close-up showing the southern portion of  Orion that contains the
OB1c and d  sub-groups of  the Orion OB1 association.    Various
clusters are marked.  The older 1c subgroup is seen directly in front
of the younger 1d sub-group.
}
\end{figure*}

The sub-cluster of stars centered on $\sigma$ Ori, located just below the eastern 
end of Orion's Belt (Figures 5 \& 6), has been assigned to OB 1b based on its
spatial proximity.   However, the 
ages of the $\sigma$ Ori stars indicate that it is considerably younger than most 
of the stars in the Belt region (Walter et al. 1997).     Because of its similar age, 
I prefer to assign the loose  cluster of 2 to 4 Myr old  stars centered on $\sigma$ Orionis 
about 1$\rm ^o$ south of $\zeta$ Ori at  the east end of the Belt (see the chapter by 
Walter et al. ) to the OB 1c sub-group.  

The Orion Nebula Cluster (ONC) in the Orion A molecular cloud and
NGC~2024 in the Orion B molecular cloud are the two largest clusters in
the youngest subgroup (known as OB1d) with ages less than 2 Myr (see
chapters by Muench et al. and O'Dell et al. for the ONC and Meyer et
al. for NGC~2024).  In addition to these large clusters, the 1d
subgroup contains a dozen smaller clusters and a background
distribution of relatively isolated stars forming in cores throughout
the Orion molecular clouds (e.g. NGC 2068 and 2071 in Orion B, see
chapter by Gibb, and L1641, see chapter by Allen \& Davis). Several
thousand (mostly low mass) members of the OB1d subgroup formed from
the ``integral shaped filament'' (Bally et al.  1987; Johnstone \&
Bally 1999) in the northern part of the Orion A molecular cloud that
contains the Orion Nebula.  About 2,000 of these stars with ages $<
10^6$ years are concentrated around the massive Trapezium stars in the
Orion Nebula itself (Hillenbrand 1997).  Hundreds more are forming in
the dense OMC2 and 3 cores north of the Orion Nebula (see chapter by
Peterson \& Megeath).  Finally, there are dozens of small, mostly
cometary clouds spread around the Orion region that form small
aggregates (see chapter by Alcala et al.).  Though the full membership
of the OB association has not been established, between 5000 and
20,000 stars are likely to have formed in the Orion region within the
last 15 Myr.  The ages and locations of Orion's various subgroups
indicate that star formation has propagated through the proto-Orion
cloud in a sequential manner.

The older subgroups in Orion lie closer to us than the younger
subgroups.  While the distances to the brighter members of the older
1a and 1b subgroups range from 320 to 400 pc (Brown et al. 1994,
1995), the Orion Nebula was thought to lie at a distance of about 400
-- 480 pc (Warren \& Hesser 1977; Genzel et al. 1981).  However,
recent radio-parallax measurements using very long baseline
interferometry indicate that the Orion Nebula is located at a distance
of 437$\pm$19 pc (Hirota et al 2007) , 389$\pm$23 pc (Sandstrom et
al. 2007), or 414$\pm$7 pc (Menten et al. 2007), see detailed
discussion in chapter by Muench et al.  The 2 to 6 Myr old OB1c
subgroup lies in front of the northern part of the Orion A cloud and
the Orion Nebula.  The absence of any obvious illumination or
reflection nebulosity on the surface of the Orion A molecular cloud
that can be attributed to massive members of the OB1c subgroup implies
that this older subgroup must lie at least 10 pc in front of the Orion
A cloud (see Figure 6).

\begin{figure*}
\center{\includegraphics[width=1.0\textwidth]{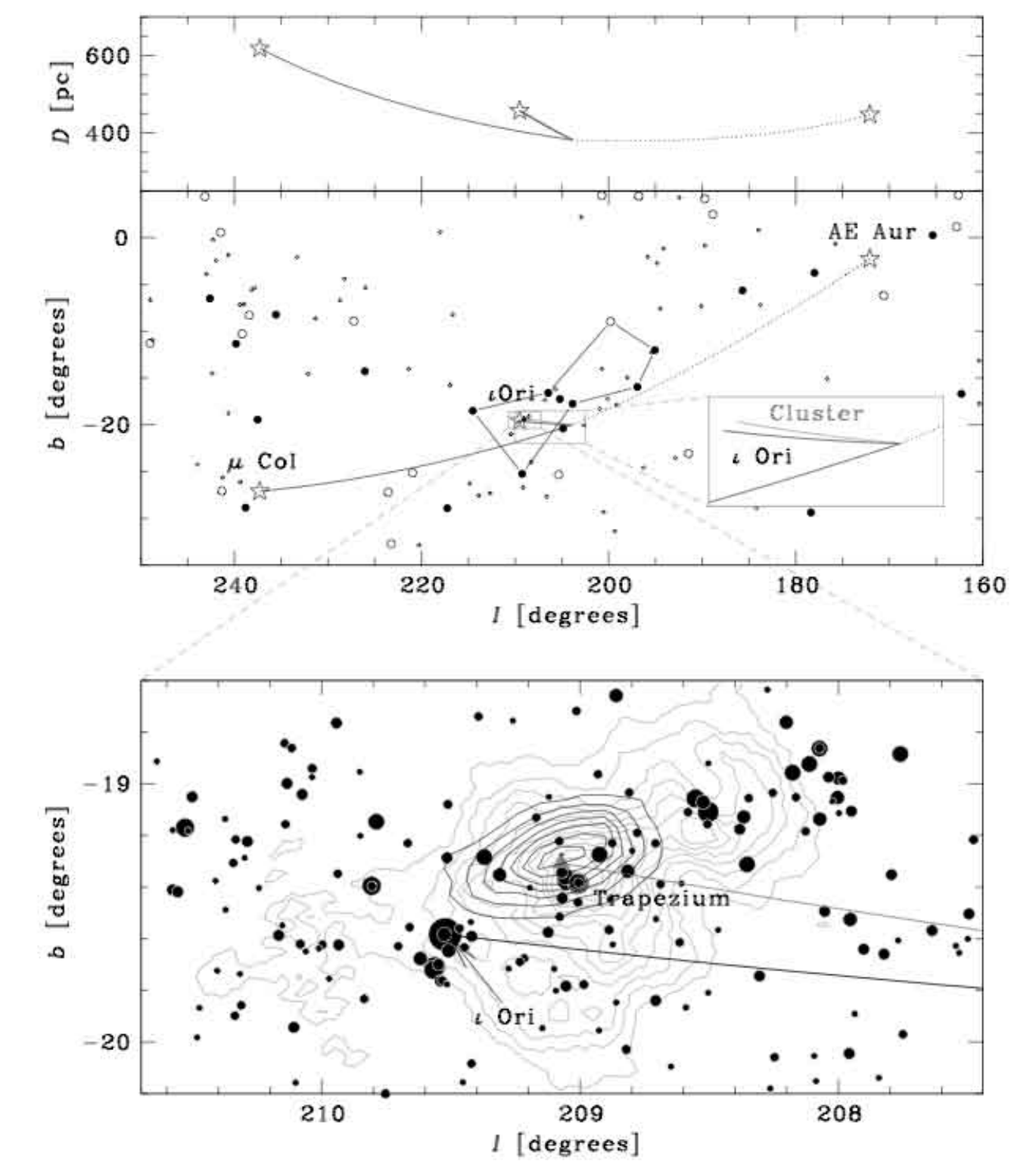}}
\caption{\small
The run-away stars AE Aur and $\mu$ Col were ejected from
the vicinity of the Orion Nebula about  2.6 Myr ago.  Proper motions
indicate that both the material that formed the ONC cluster in the
Orion Nebula and the cluster NGC 1980 were co-located with
point of origin of these runaway stars.     The Orion Nebula and its
stars had not yet formed.  But  the cluster
NGC 1980  contains the colliding-wind X-ray binary $\iota$ Ori. 
Gualandris et al. (2004) claim this binary, and the ejection of the two
run-away stars, formed when a four-member non-hierarchical massive 
star system in NGC 1980 decayed, forming the binary, and ejecting the
two least massive members.   Taken from Hoogerwerf et al. (2004).
}
\end{figure*}

It has been known for decades that run-away stars are common
among O stars,  rare among B stars, and virtually non-existent for later spectral 
types (Gies \& Bolton  1986; Gies 1987).    Orion has been the source of 
several well known run-away 
stars, including the 150 \kms\ run-away star AE Auriga,   and the 117 \kms\ 
$\mu$ Columbae which is moving exactly in the opposite direction
(Blauuw 1991).  Hoogerwerf, de Bruijne, and  de Zeeuw (2001) used new 
Hipparcos proper motion data to show that these two stars, and the colliding
wind X-ray binary $\iota$ Ori were at the same location in the sky 2.6 $\pm$ 0.05
Myr ago.    Gualandris, Portegies-Zwart, and Eggleton (2004) argue that
the two run-away stars and $\iota$ Ori suffered a four-body interaction in which
two binaries in the same cluster underwent an exchange.  The two most-massive
members became the tight $\iota$ Ori binary;  the gravitational energy released
kicked the two less massive stars out of the region at high velocity.  Figure 7 shows 
current configuration of these stars on a wide-field star-chart, along with the stellar
motions traced back over time.  

Interestingly, the proper motion of the Orion Nebula  Cluster would place it at the
location of the four-body interaction 2.6 Myr ago.  The presence of some older 
stars  in  the Orion Nebula Cluster (ONC) indicates that  some star formation did 
occur in this region.   However, the number of older stars indicate
that star formation in the gas that created the ONC was much slower 2.6 Myr ago.
Evidently the star formation rate in the ONC has been accelerating over time,
culminating in the relatively recent formation of the Trapezium group of massive
stars.  Massive star formation continues to this day in the dense molecular ridge
immediately behind the Orion Nebula.

The star $\iota$ Ori is embedded in an  older, relatively dispersed
cluster,  NGC 1980,  currently located at the  southern end of  Orion's Sword.
NGC 1980 has been associated with the 1c sub-group of the Orion OB association
located directly in front of the Orion Nebula and associated Orion A cloud.  
This dispersed cluster is located in lower part of Figure 2 in the Chapter by 
Muench et al. in this volume and was designated as group V by Parenago (1954)
and as Group 5 by Walker (1969) - see Table 2 in the Chapter by Muench.  The Orion
OB1c sub-group also contains the cluster NGC 1981 located north of NGC 1977
and the Orion Nebula (see Figure 6).    
Assuming that  NGC 1980 shares its motion through space with $\iota$ Ori, 
this cluster also would have been located in the same region as the material
that later formed the Orion Nebula and rich cluster.    Since $\iota$ is 
currently the most massive member of NGC 1980, this cluster is the most likely 
parent of the the  two run-away stars AU Aur and $\mu$ Col.   That the material 
that turned into the Orion Nebula and the ONC was apparently close to NGC 
1980 several million years ago, suggests the intriguing possibility that the 
formation of the ONC and the  Nebula may  have been triggered by the 
older NGC 1980 cluster.    If this hypothesis is correct, then the older
stars in the ONC may in fact be outlying members of NCG 1980 and the 
Orion OB1c subgroup.

The locations and ages of stellar groups in the Orion region indicate that
star formation can propagate through a cloud in a non-linear fashion.    As first
generation stars trigger the birth of subsequent generations, stellar groups with
similar ages may  be formed is spatially disconnected regions of the sky.   In 
Orion, the 1a subgroup appears to be the first to have formed.  It massive stars
may have induced the birth of the 25 Ori and 1b subgroups.   Subsequent 
triggering may have induced further star formation in Orion's Sword to the south, 
$\sigma$ Ori to the southeast, and  possibly $\lambda$ Ori to the north.   Within
the last few Myr, star formation propagated into the Integral Shaped Filament
in the Orion A molecular cloud to form the Orion Nebula, M43, OMC2, OMC3, and 
NGC 1977, the "traditional members of the 1d sub-group located directly behind 1c.  
Recent star formation has also ignited east of Orion 1b and the $\sigma$ Ori
group to give rise to NGC 2023, NGC 2024, NGC 2068, and NGC 2071.  Additionally,
small stellar groups are forming from dozens of widely scattered cometary cloud
strewn throughout the interior of the Orion super-bubble.    

The lifetimes  of the various molecular clouds in Orion remain unclear.  
Did the entire OB association form from a pre-existing giant cloud, or are the
clouds formed rapidly by compression of the surrounding ISM as the super-bubble
created by Orion's massive stars sweeps-up and compresses the medium?  The
structure and kinematics of the Orion A and B clouds and well as the dozens of smaller
clouds in the region suggests that a "sweep-up, compress, and trigger" scenario may
be at work.     What roles did the massive run-away stars play in shaping the Orion
super-bubble and in triggering cloud and star formation in distant regions?  
These issues will be addressed in the next section.

\begin{figure*}
\center{\includegraphics[width=1.2\textwidth]{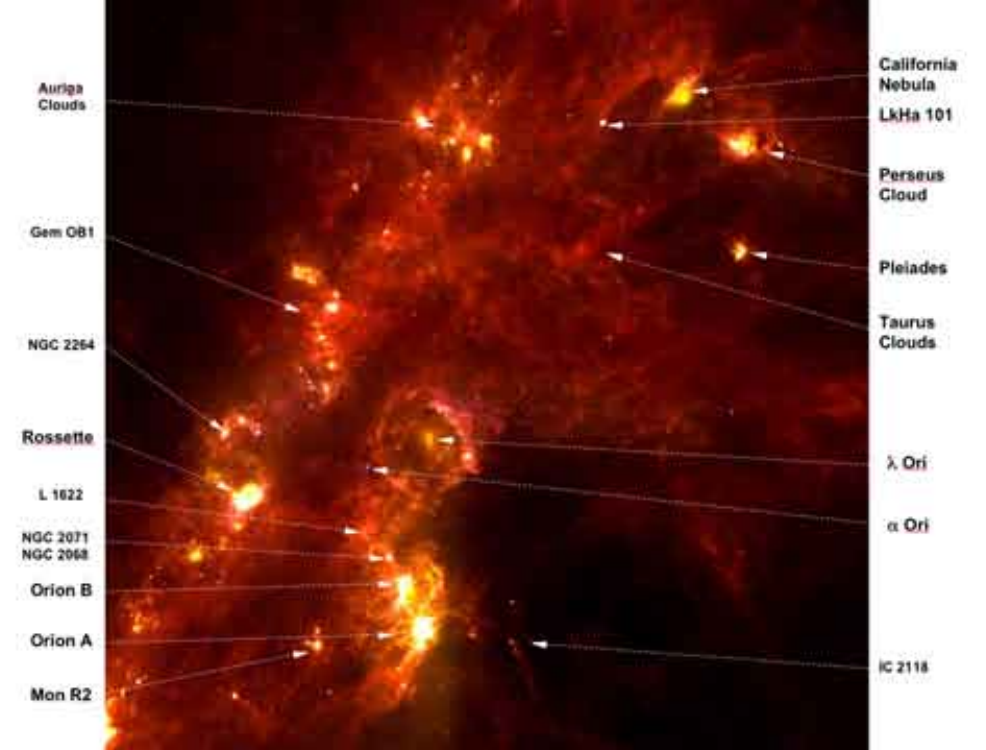}}
\caption{\small
A wide field 100 $\mu$ IRAS view of the Orion regions and its
surroundings.  Various star forming regions are marked.
}
\end{figure*}

\section{The Superbubble and shell:  Orion's Cloak}

Massive stars inject energy into the ISM through their Lyman continuum 
radiation, stellar winds, and supernova (SN) explosions.  If one assumes a
standard IMF, then the estimated population of young stars in Orion
implies that between 30--100 stars more massive than 8~M$_{\odot}$
have formed in this region in the past 12 Myr.  Many of these massive 
stars have evolved off the main sequence and exploded
during the past 10 Myr.   Using the age/mass relationship 
$\rm \tau (M) ~=~ k M^{-\beta}$, with $\beta ~=~ 1.6 \pm 0.15$ in the 
mass range 8 to 80 M$_{\odot}$ (Shull \& Saken 1995), stars in the 
1a subgroup more massive than about 13~M$_{\odot}$ have exploded. 
For the 1b and 1c subgroups, a median age of 6 Myr implies 
that all stars more massive than about 20~M$_{\odot}$ have exploded.
Thus, there have been 10 to 20 SN explosions in the Orion region
during the last 12 Myr.  The released kinetic energy ($> 10^{52}$~ergs) 
has formed a large bubble of X-ray emitting gas whose expansion
has swept-up a massive shell of gas and dust, the Orion/Eridanus
superbubble.    

Figure 8 shows a wide-anlge view of the Orion region using
the 100 $\mu$m IRAS data which shows the distribution of warm dust.
The Winter Milky Way stretches diagonally from the lower-left corner to
the middle of the top boundary of the image; various well known star-forming
complexes are marked.   The brightest 100 $\mu$m emission traces
the warm dust associated with the young HII regions in 
Orion such as the Orion Nebula and NGC 1977 in Orion A,  and NGC 2024 in
orion B.  The spectacular $\lambda$-Ori ring is located just below and left
of center.   The dark region west of Orion A and B that dominates the
lower right of Figure 8 traces the interior of the cavity excavated in the 
ISM by Orion/Eridanus superbubble.   

Over 100 years ago, Barnard discovered a crescent of H$\alpha$ emission 
that wraps around the eastern portion of Orion (Figure 9).  Sivan (1974) 
and Reynolds \& Ogden (1979) demonstrated that Barnard's Loop is the 
brightest part of a giant bubble of H$\alpha$ emission that extends 
40$\rm ^o$ west into Eridanus.  The bubble subtends $\rm 20^o ~\times ~40^o$, 
which at a distance of 400 pc corresponds to about 140 by 300 pc 
(Figures 9 and 10).  The bubble appears to be blowing out of the Galactic
plane since the OB association and its parent clouds lie 
about 15$\rm ^o$ to 20$\rm ^o$ (100 to 150 pc) below the plane.   
As a result, the bubble is expanding into a steep density (and pressure) 
gradient with a mean velocity of order 10 to 20 $\rm km~s^{-1}$ and may 
be evolving towards blow-out from the Galactic plane (Mac Low \& McCray 1988).
The \Ha\ bubble  has a mass of about 
$\rm M(HII )~=~7 \times 10^4~M_{\odot} d^2_{380}$  
($\rm d_{380}$ is the distance in units of 380 pc) and a 
kinetic energy of at least $\rm 1.7 \times 10^{50} d^2_{380}$ ergs
(Reynolds \& Ogden 1979; Cowie et al. 1979; Burrows et al. 1993).
The temperature in the bubble interior has been estimated to be 
$\rm 1~to~5\times10^5$ K from the excitation of UV lines and
the presence of soft X-ray emission. 
It has also been detected in the 1.8 MeV $\gamma$-ray line
of the short-lives radioactive species $^{26}$Al,  possibly indicating extensive
recent pollution of the bubble interior by supernovae  (Figure 11
and Diehl et al. 2004).  Cowie et al. (1979) dubbed the hot
gas in the interior of the Orion/Eridanus superbubble ``Orion's Cloak''.

The walls of the Orion superbubble are visible in the far infrared 
(Brown et al. 1995) and in 21~cm \Hi\ emission (Heiles 1976; Green 1991; 
Green \& Padman 1993).  Both tracers exhibit filamentary
structure with most of the emission coming from a region located well
outside the H$\alpha$ shell.  The \Hi\ mass of the shell 
is about $\rm 2.3 \times 10^5d^2_{380}~M_{\odot}$ 
and its kinetic energy is about $\rm 3 \times 10^{51}d^2_{380}$ ergs.   
Absorption measurements towards stars with known distances indicate
that the near-wall of the Eridanus Loop may be as close as 180 pc from the 
Sun.   Figure 12 show a possible geometry of the bubble and its walls.  
The proximity of this portion of the Orion/Eridanus Bubble and the presence of 
hot  plasma  may be an indication that a supernova
exploded about midway between Orion and the Sun within the last few Myr.

\begin{figure*}
\center{\includegraphics[width=0.9\textwidth]{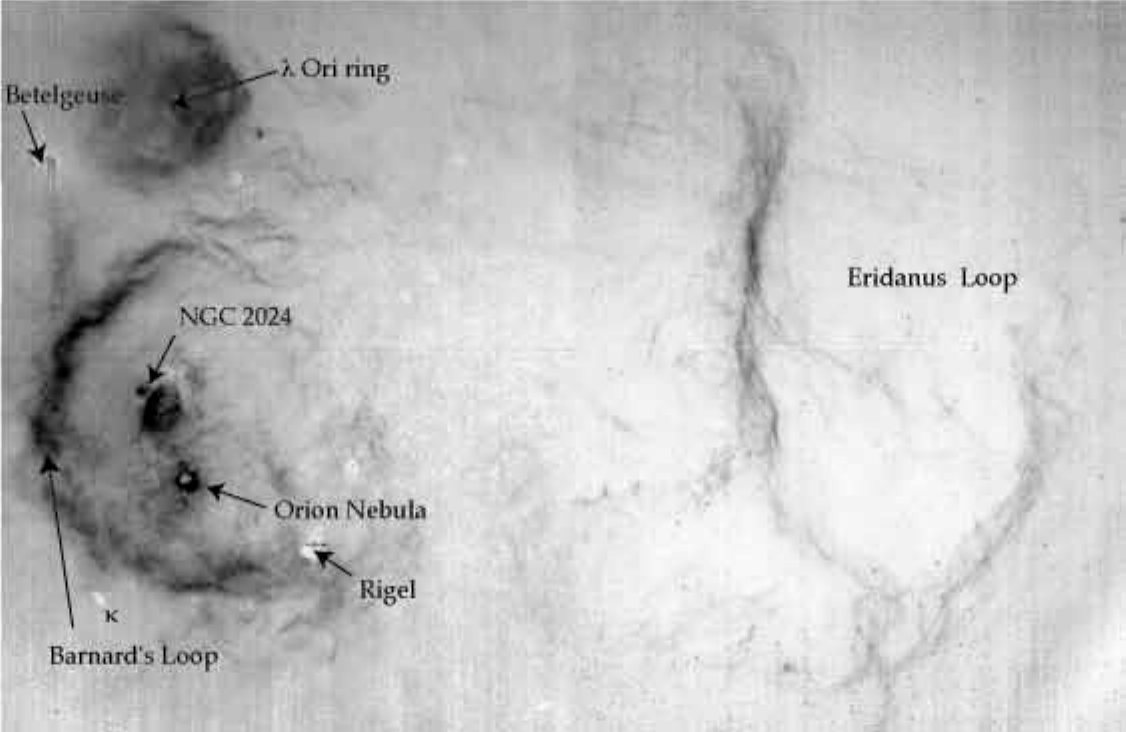}}
\caption{\small
A continuum subtracted log intensity scale image of the 
Orion/Eridanus Bubble in H$\alpha$ showing a 
$30^{\rm o} \times 47^{\rm o}$ field of view.
}
\end{figure*}

\begin{figure*}
\center{\includegraphics[width=1.0\textwidth]{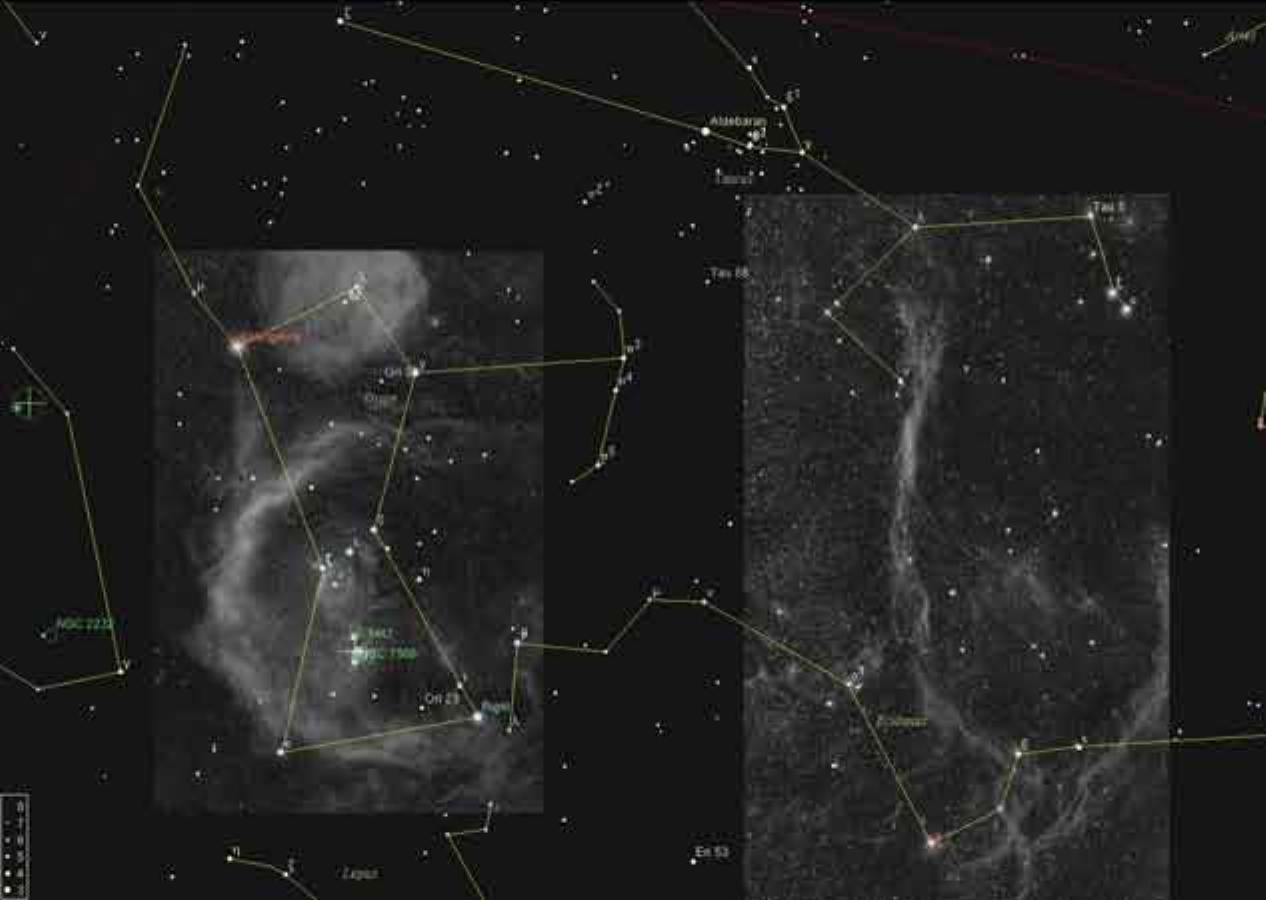}}
\caption{\small
Wide-field H$\alpha$ images of the Orion / Eridanus bubble
overlaid on a star  chart.  The image is shown in equatorial
coordinates. Courtesy A. Melinger. 
}
\end{figure*}

\begin{figure*}
\center{\includegraphics[width=0.8\textwidth]{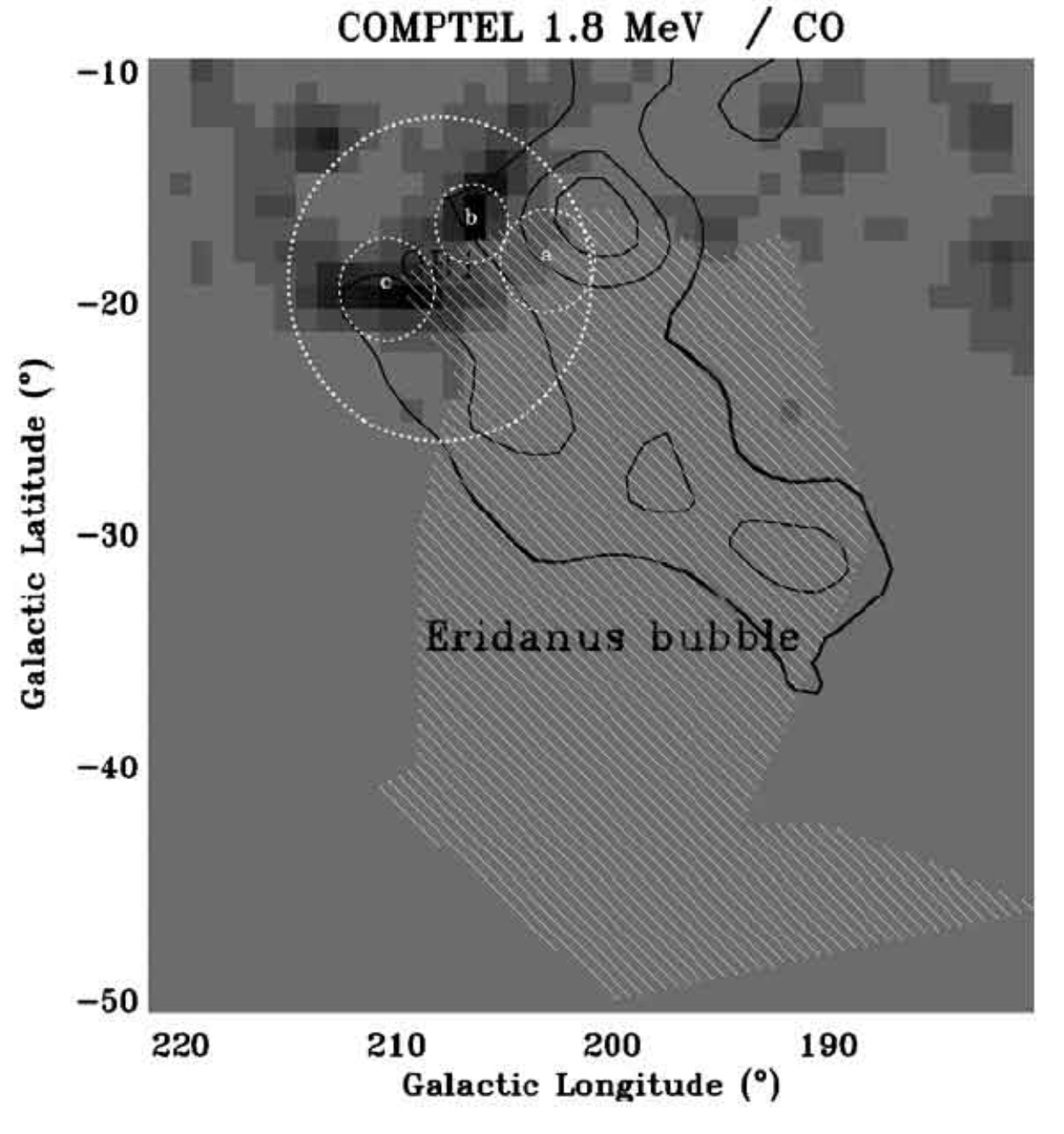}}
\caption{\small
Wide-field view of the Orion / Eridanus bubble
showing contours of 1.8 MeV gamma-ray emission from
decays of $^{26}$Al measured by COMPTEL (black contours), 
superimposed on a very low resolution CO map from Dame et al. (2001) 
shown in  grey-scale.   The white hatched area shows the location
of the Orion-Eridanus cavity.  The locations of the Orion OB1a, 1b, and 1c 
subgroups of the Orion OB association are shown in white.
Taken from Diehl et al. (2004).
}
\end{figure*}

\begin{figure*}
\center{\includegraphics[width=0.8\textwidth]{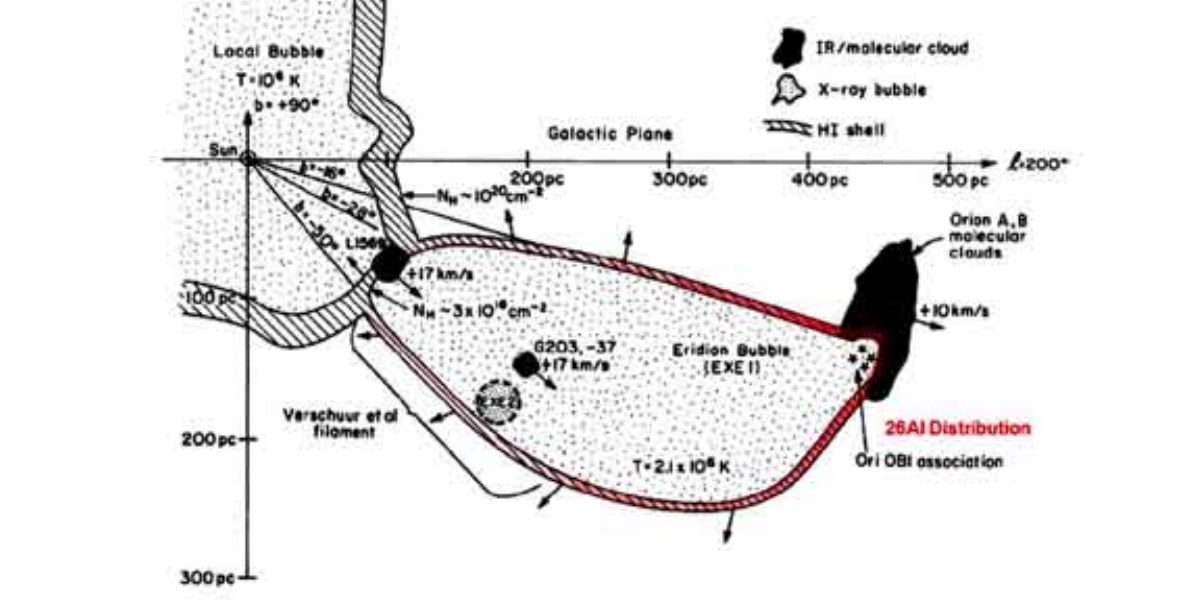}}
\caption{\small
A cartoon showing the geometry and location
of the Orion / Eridanus bubble relative to the Sun
and the Orion molecular clouds.
Taken from Diehl et al. (2004).
}
\end{figure*}

Orion contains two $\rm 10^5~M_{\odot}$ giant molecular
clouds; the Orion A cloud located behind Orion's Sword in the
southern portion of the constellation, and the Orion B cloud that
lies east of Orion's Belt.  Both GMCs are located in the projected 
interior of the Orion/Eridanus superbubble (Figure 8 \& 9) 
and appear to have been shaped by energy release from the 
Orion OB association (Maddalena et al. 1986; Bally et al. 1987; 1991a, b).

Orion A is cometary in appearance with a compact ridge of dense 
gas at the northern end (the ``integral shaped filament'') 
and a lower density and wider tail that 
extends directly away from the centroid of the OB association.  
This cloud exhibits a large scale velocity gradient along its length;  
the northern part of the cloud 
is centered at $\rm v_{\rm lsr} ~=~ 11~km~s^{-1}$ while the 
southern part near $\kappa$ Orionis has components with
$\rm v_{\rm lsr} ~=~ 2~km~s^{-1}$.  Bally et al. (1987) interpreted
the smaller transverse extent and relatively high mean density of
the gas in the north and the large scale velocity gradient as evidence
for compression and acceleration by the OB association.  Thus, star 
formation in the northern part of Orion A may be an example of 
triggering (Elmegreen 1997b).    Orion A may be forming at its
southeast end as the ISM is swept-up and compressed by the advance of 
Barnard's Loop; it may be destroyed by massive stars at its northwest end.

The Orion B cloud contains a high density ridge at its western 
end where the embedded clusters NGC~2024 and NGC~2023 and the
Horsehead Nebula are located.  A network of dust and CO 
filaments trails away from the OB association towards the east.
The northern part of Orion B contains the NGC~2068 and NGC~2071
clusters that are embedded in dense cores at the southwestern ends
of cometary clouds of CO emission.  These clouds also trail off in 
a direction pointing away from the OB association.

Dozens of smaller cometary clouds scattered throughout the
interior of the Orion superbubble also have tails that point 
directly away from  the Orion OB association.  While a few of these 
clouds have velocities similar to the Orion A and B GMCs, 
the ensemble of small clouds in Orion has a large velocity
dispersion that extends over a range $>$ 20 \kms .  While this
is far higher than the velocity dispersion of the Orion A and B
clouds, it is smaller than the range of \Ha\ radial velocities 
associated with the \Ha\ shell.   The CO velocities of the cometary 
clouds may indicate that the more massive fragments have suffered less 
acceleration while the smaller fragments either suffered larger 
accelerations or condensed from the tenuous \Hi\ gas comprising the 
expanding Orion/Eridanus superbubble.

\begin{figure}
\includegraphics[width=0.9\textwidth]{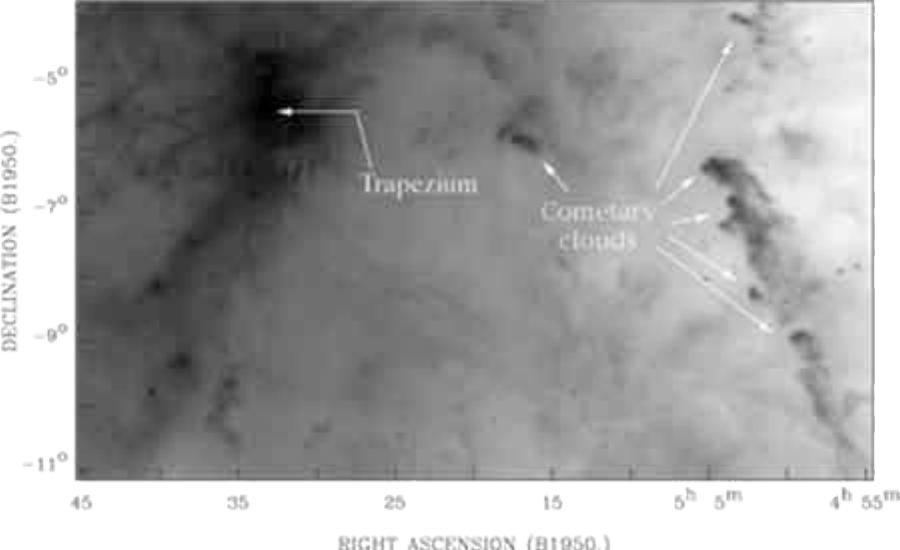}
\caption{An IRAS 100~$\mu$m view (log scaling)
          of a family of cometary clouds
          located about 5$\rm ^o$ west of the Orion A cloud.
          The dark region extending southeast of Trapezium
          is the Orion A cloud.  The Cometary cloud above the
          center of the image is L1634.  IC 2118 consists of
          the  cometary clouds 
          near the right edge of the image.}
\label{fig-13}
\end{figure}

\begin{figure}
\center{\includegraphics[width=0.6\textwidth]{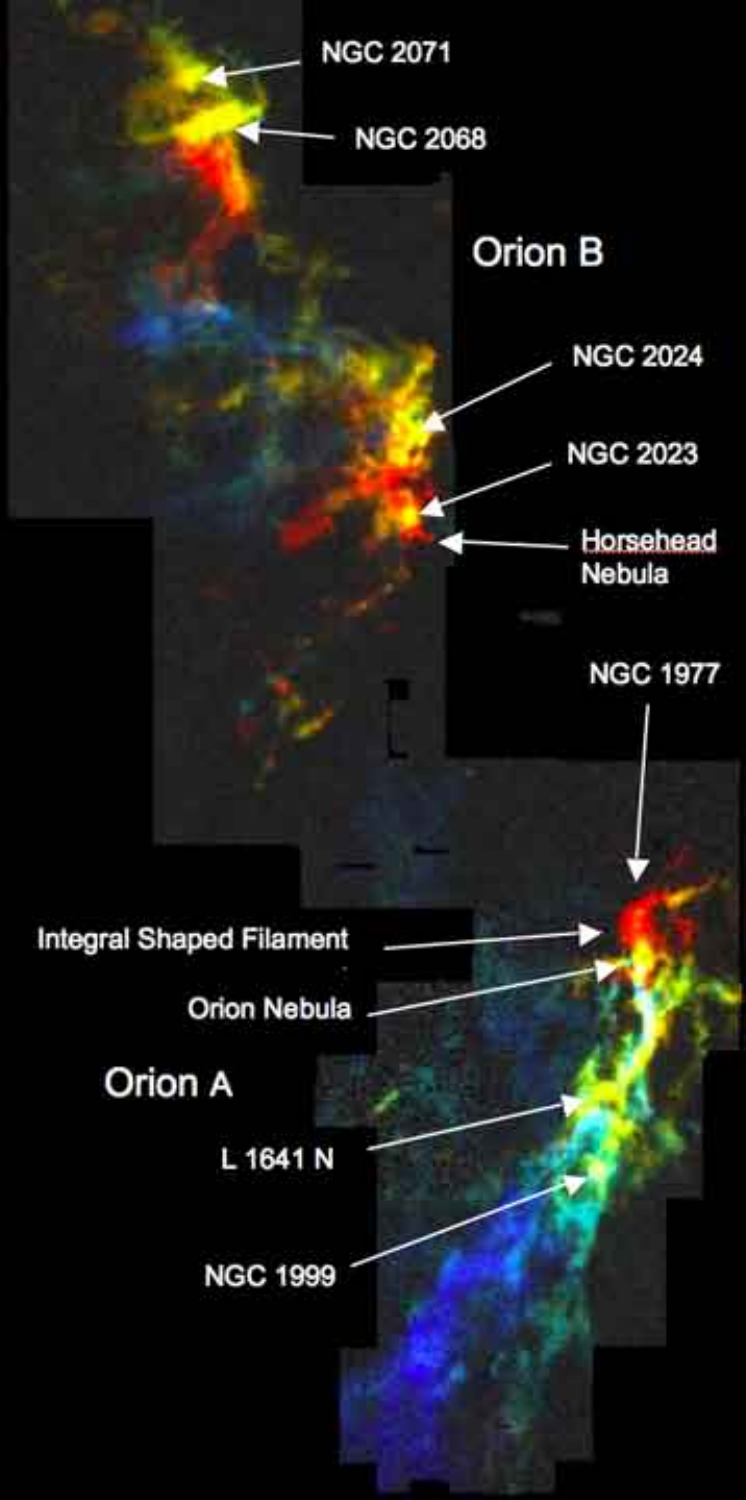}}
\caption{A 110 GHz $^{13}$CO J=1-0 image showing the Orion A and B clouds.  The colors 
represent Doppler shifts with 
blue corresponding to  V$_{LSR}$ = 0 to 5 km s$^{-1}$,  
green corresponding to  V$_{LSR}$ = 5 to 10 km s$^{-1}$, 
and blue corresponding to  V$_{LSR}$ = 10 to 15 km s$^{-1}$.
Data were obtained during the mid 1980s with the 7 meter
radio telescope located on Crawford Hill in Holmdel, NJ. (see Bally et al. 1987). }
\label{fig-14}
\end{figure}

\begin{figure}
\includegraphics[width=0.9\textwidth]{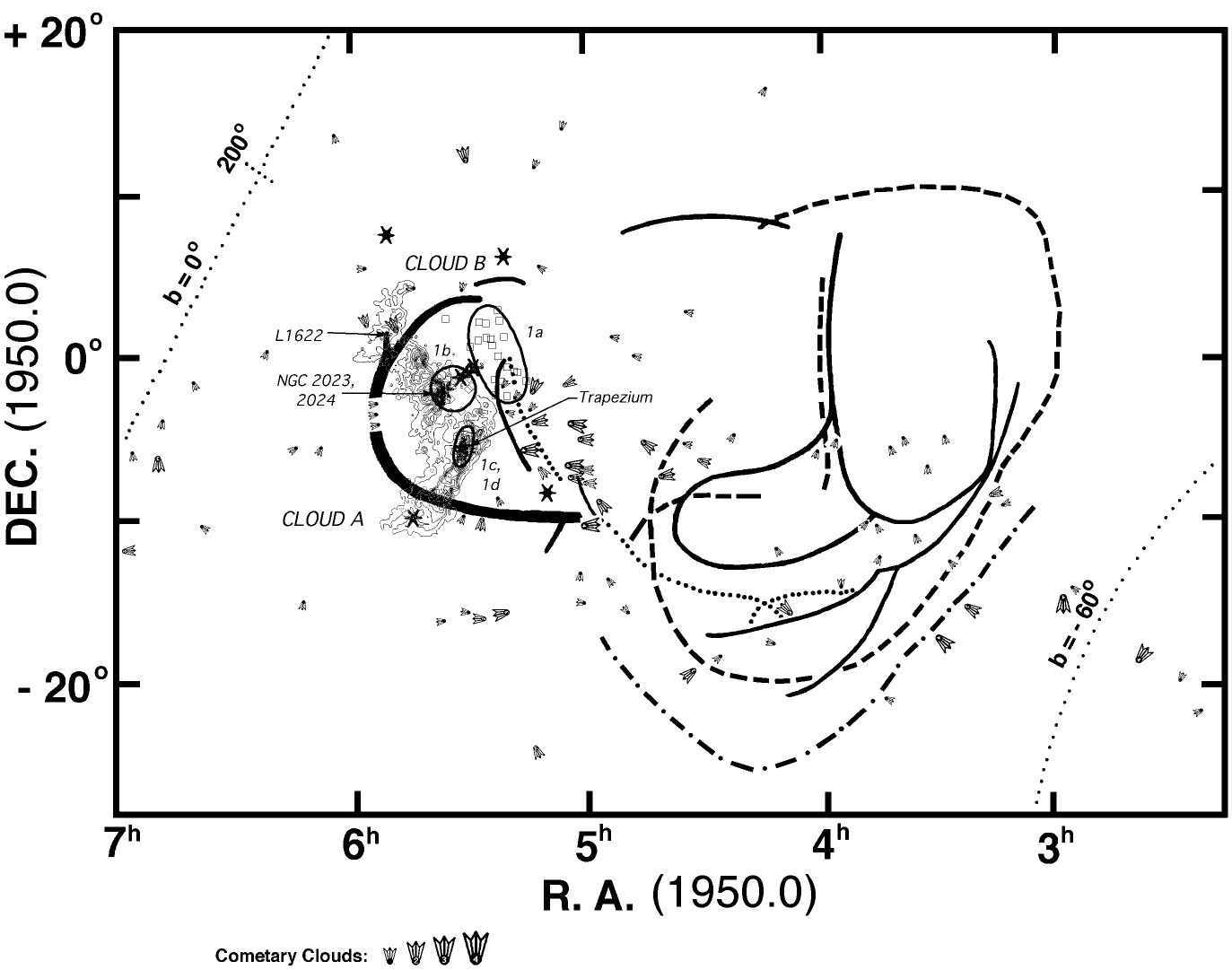}
\caption{A cartoon showing the locations of cometary clouds detected
using IRAS 100 $\mu$m emission from warm dust.  The faint
solid contours show the location of the Orion A and B clouds shown in
detail in Figure 14.  The thick
solid lines show the location of the brightest H$\alpha$ emission
from Figures 9 and 10.  The dashed and dot-dashed lines show the
location of ridges of 21 cm HI  emission thought to be associated with
the Orion-Eridanus bubble.
}
\label{fig-15}
\end{figure}

Figure 13 shows several dozen cometary clouds located about 5$\rm ^o$
west of the Orion A cloud.  The large cloud on the right side of this image
is the IC 2118  cloud complex lit-up by Rigel.   The large-scale
structure of IC 2118 is a giant cometary structure 
with a major axis pointing back towards the older (10 Myr) Orion OB1a
subgroup that lies about 5$\rm ^o$ to 8$\rm ^o$ to the northeast.
Both  the Orion A cloud and IC 2118 point to the OB1a subgroup.
However, individual sub-condensations, especially on the eastern
side of IC 2118, have axes of symmetry that point towards the northern
portion of Orion A where the foreground OB1c and youngest OB1d
subgroups are located.   It is likely that initially, this cloud complex was
shaped by UV radiation emitted by the older OB1a and 1b subgroups.  
However, within the last few Myr, UV radiation from the 
the OB1c subgroup near the Sword may have exerted a stronger influence.
The influence of this younger group is especially evident in the shapes of the
L1515/L1516 cloud located north of IC 2118, and in the L1634 cloud located
about 3\deg\ due west of Orion's Sword.   The expanding wind of tenuous 
plasma that is responsible for inflating the Orion-Eridanus bubble may also 
have contributed to the shaping of the cometary clouds in its interior. 

The distribution of young stars in these cometary clouds indicate that star formation
has propagated from east-to-west.   Unobscured classical and weak-line T 
Tauri stars are found predominantly east of the ionization fronts that line the 
eastern rims of these clouds.  Younger, embedded YSOs that are seen mostly at
infrared wavelengths and which drive outflows, tend to be located to the west
of these ionization fronts.

Figure 14 shows a $^{13}$CO image of the Orion A and B clouds illustrating
the cometary morphology and velocity gradient in Orion A, and the 
complex structure of Orion B.
Figure 15 shows a cartoon illustrating the relationships between 
the CO clouds, the Orion/Eridanus bubble as traced by H$\alpha$, the 
21 cm HI emission, and ridges
of warm dust emission, and the cometary clouds identified in the IRAS
infrared images.

The structure of the interstellar medium in the Orion/Eridanus bubble 
provides evidence that the energy released by high mass stars has 
profoundly altered the morphology, structure, and kinematics of the 
gas in this region.  Over the past 5  to 10 Myr, a 20$\rm ^o$ by 40$\rm ^o$ 
diameter bubble has been inflated by the OB association.  H$\alpha$ 
emission traces the location of the current ionization
front of the Orion complex.  This low density gas is expanding with 
a mean velocity of about 10 to 60 \kms\ towards high Galactic latitudes 
and towards us.  This gas can be seen in absorption against Orion's
bright stars, especially in the UV.   A tenuous and hot plasma, visible in soft X-rays 
and in UV resonance lines, fills the bubble interior and the the 
near-side of the bubble may be as close as 180 pc from the Sun. 
The western portion of the Orion / Eridanus bubble has been
detected in the 1.8 MeV gamma-ray line produced by decays of the
short-lived ($t_{1/2} \approx$ 0.7 Myr) species $^{26}$Al (Diehl et al.  2004
-- see Figure 11).  Figure 12 shows the apparent geometry of the Orion
/ Eridanus bubble.

A larger shell of 21 cm \Hi\ emission and dust traced by IRAS and COBE 
data lies outside the H$\alpha$ emission region.  The remnants of the 
giant molecular cloud from which the older association stars formed 
have been overrun by the expanding superbubble and were 
accelerated and compressed.  The parts of this cloud that lie close
to the OB association have higher radial velocities than the gas located
farther out.  This trend may indicate that the molecular cloud
lies behind the superbubble and the majority of the Orion OB association's 
stars.  Apparently star formation occurred preferentially on the 
near side of the cloud and the resulting release of energy has pushed the 
surviving cloud remnants away from us.  In addition to the Orion A and B
GMCs, nearly a hundred smaller clouds are strewn throughout the 
interior of the Orion/Eridanus bubble and most of these cometary clouds point 
back towards the Orion OB association.  All are visible in the IRAS data, most 
have bright skins of H$\alpha$ emission, and many emit in the CO lines.  
The cometary CO clouds have slower velocities away from OB 
association than the \Ha\ or 21~cm \Hi\ shell, but larger speeds than the
Orion A and B clouds.   While some small clouds may have condensed
from the expanding supershell as it evolves towards blow-out at high
Galactic latitudes and becomes unstable to Rayleigh-Taylor instabilities, 
the larger clouds are probably accelerated remnants of the proto-Orion GMC.

\section{The Origin of Orion's Brightest Stars}

Where did Orion's luminaries, Betelgeuse ($\alpha$ Ori; M2Iab) and
Rigel ($\beta$ Ori; B8Iab), and the 2nd magnitude Saiph ($\kappa$ Ori;
B0Iab) originate?.  The fourth star in Orion's rectangular shape,
Bellatrix ($\gamma$ Ori; B2III) in the northwest corner of Orion, is
located only 75 pc from the Sun, and is presumably unrelated to Orion.

Betelgeuse has been estimated to lie 131 $\pm$30 pc from the Sun based
on Hipparcos measurements (Perryman et al. 1997).  However, it is a
supergiant with a pulsating atmosphere which makes parallax estimation
difficult, especially at visual wavelengths below the peak of the
Planck function where slight photospheric temperature variations can
shift the photo-center.  A recent VLA-based distance estimate which is
based on the photo-center in the Rayleigh-Jeans limit, places it at a
greater distance of 197 $\pm$45 pc from the Sun (Harper, Brown, \&
Guinan 2008), about 200 pc in front of the Orion Nebula.

Betelgeuse is a moderate-velocity run-away star with a proper motion of 
($\mu _{\alpha \rm{cos} (\delta)}$, $\mu_{\delta}$) =  
[23.98 $\pm$1, 10.07 $\pm$1.15] mas / yr (Boboltz et al. 2007) which indicates
motion towards the northeast at $PA$ = 68\deg, directly towards the
Galactic plane.    An infrared bow shock
precedes  the motion of this star (Noriega-Crespo et al. 1997).
The heliocentric radial velocity of Betelgeuse is 
$V_r$  = 21.0 km~s$^{-1}$.  This is 7 km~s$^{-1}$ lower than the radial velocity
of the stars in the Orion Nebula, 10 km~s$^{-1}$ lower than the majority of stars
in the Orion OB1a subgroup, and comparable to the 25 Ori stellar aggregate
in the 1a subgroup.    The new distance led Harper et al. (2008) to re-determine
the current mass (17 M$_{\odot}$) and birth mass (20 M$_{\odot}$) of Betelgeuse, 
implying that this star had a main-sequence lifetime of about 10 Myr, comparable to
the ages of stars in the OB1a subgroup.  However, the 3D motion of Betelgeuse
propagated back in time is inconsistent with a direct ejection from this
group.   

From the proper motions of $\alpha$ Ori ([24,10] mas/yr = 0.026$"$/yr)
and the angular separation of $\alpha$ Ori and 25 Ori, the B1 star in
the 25 Ori group in Ori OB1a that is about 9.4 degrees from Betelgeuse
on the sky, these two stars were closest to each other about 1.3 Myr
ago (assuming that the proper motion of 25 Ori is negligible).  The
radial velocity difference between Betelgeuse and the 1a subgroup of
Ori OB1 (excluding the 25 Ori group, which has a radial velocity
identical to Betelgeuse) is 10 km/s.  To cover the 130 pc difference
between the distances of Betelgeuse and the 1a subgroup would require
13 Myr - very different from the 1.3 Myr obtained from the proper
motions alone.  Betelgeuse and the OB1a subgroup could not have been
at the same location at any time.

The main sequence life-time of 10 Myr places a severe constraint on
the point of origin of Betelgeuse.  If its current proper motion was
imparted at birth, Betelgeuse would have been born far above the
Galactic plane.  Thus, its velocity can not have originated at birth.
Either, Betelgeuse was born in an unknown group located southwest of
the Orion OB1 subgroups, or if it was born in OB1a, as suggested by
its age, it must have gotten two velocity kicks; an initial kick that
moved it from its birth-place in OB 1a to within about 200 pc from the
Sun, and a second kick about 1 to 2 Myr ago that generated its current
proper motion.

$\beta$ Ori (Rigel) has a distance of about 245 $\pm$45pc, similar to
Betelgeuse and the IC 2118 cloud and its population of T Tauri stars
that is externally illuminated by Rigel.  The radial velocity
(Kharchenko et al. 2007) is $V_r$ = 18.8 to 20.3 km~s$^{-1}$ and
proper motion is [1.87,--0.56] mas/yr (Hog et al. 2000).  Thus, there
is evidence for foreground star formation about 200 to 300 pc from us,
well in front of Ori OB1. Could Betelgeuse have been ejected from this
group, perhaps by a companion that exploded?

Alternatively, the OB1a subgroup might have dynamically ejected a
massive binary roughly toward the Sun about 10 Myr ago. At V = 10 to
15 km/s, it would have moved about 100 to 150 pc towards us.  If this
binary consisted of Betelgeuse and a more massive star, the massive
companion could have exploded in a supernova event about 1 to 2 Myr
ago - this constraint arising from the requirement that Betelgeuse be
within about 15 degrees of its current location when it got its
present motion. The explosion would have converted the orbital motion
of the Betelgeuse into linear motion, setting it on its current path.
The site of such a supernova explosion would have occurred somewhere
between Orion OB1a and the Eridanus Loop.  A supernova at this
location may explain several aspects of the Orion / Eridanus
superbubble such as the extension toward the Sun shown in Figure 12,
the location of the near wall (the Eridanus Loop) at a distance of
only 180 pc from the Sun, the presence of soft X-ray and 1.8 MeV
$\gamma$-rays from the decay of live $^{26}$Al.

$\kappa$ Ori (Saiph), the bright star located at the southeastern
corner of Orion, also has a distance of about 220 $\pm$45 pc, and a
radial velocity $V_r$ = 21 km~s$^{-1}$ similar to Rigel.  Thus, this
star also provides further evidence for recent star formation in front
of Orion.  There is some circumstantial evidence that this star is
illuminating the southern end of the Orion A cloud.  Figure 8 shows a
100 $\mu$m IRAS view of the Orion region.  A ring of dust is seen to
wrap around $\kappa$ Ori, (located near the bottom of Figure 8).  CO
emission associated with this ring blends smoothly into the southern
end of the Orion A cloud, raising the possibility that the extreme
southern portion of L1641 is much closer to the Sun than the Orion
Nebula.  Perhaps the L1641 portion of the Orion A cloud is a long
finger that is elongated along our line-of-sight.

\section{Lessons Learned from Orion}

What have studies of Orion taught us?  Observations of Orion have
yielded the discovery of the first IR-nebula (the Ney-Allen nebula),
the first IR-only star (the Becklin-Neugebauer - BN object), the first
OH and H$_2$O masers, the discovery of interstellar CO and many other
molecules, the recognition that stars form from giant molecular
clouds, the first protostellar outflow, and the first direct visual
wavelength images of circumstellar disks (proplyds).

Much of our current understanding of star formation has emerged from
studies of Orion:

$\bullet$ Molecular clouds are filamentary, chaotic, and exhibit
supersonic internal turbulent motions.  It remains unclear if on a
large (GMC) scale they are bound by self-gravity and long-lived
(survive for many crossing times), or dominated by motions imparted by
surrounding flows and short-lived (about a crossing time).

$\bullet$ Stars form from dense cores embedded in molecular clouds.
Most star forming cores are located on the sides of clouds that face
the center of the OB association.  Cores fragment into sub-cores to
form groups containing from a few to thousands of stars.  Star
formation propagates through the complex at nearly the sound speed in
ionized gas.  In Orion, the 10 Myr old sub-groups are separated from
the currently forming sub-groups by about 50 pc.

$\bullet$ Most forming stars are surrounded by accretion disks and
power collimated winds and jets that inject energy and momentum into
the host cloud.  Some massive stars power explosive, poorly collimated
outflows (e.g. OMC1).

$\bullet$ The initial mass function is universal, independent of the
degree of clustering or other environmental properties.

$\bullet$ Most stars form in short-lived, transient clusters that
dissolve in a few Myr to populate the field.  The clusters and the
cores from which they form exhibit a hierarchy of structure.  OB
associations consist of sub-groups, clusters, and sub-clusters with no
preferred scale.

$\bullet$ Massive stars sculpt the ISM in their vicinity.  They create
superbubbles hundreds of parsecs in diameter and their energy and
momentum injection (UV, winds, SN) dominates the evolution of star
forming regions.

$\bullet$ Massive stars form in the densest and most massive clusters,
and appear to be born near the cluster center.  The Trapezium in the
Orion Nebula is a mass-segregated sub-cluster.

$\bullet$ Stellar multiplicity increases with mass.  Run-away stars
are common among massive stars.  While some massive stars were ejected
when a more massive companion exploded, others were ejected by
dynamical decay of dense sub-clusters soon after formation.

$\bullet$ The formation time-scale for individual stars is about
$10^5$ years, for clusters about $10^6$ years, and for entire OB
associations about $10^7$ years.

$\bullet$ The star formation efficiency, averaged over an entire OB
association, is about $\eta ~ \approx\ $1 to 5\%.

\vspace{0.5cm}

{\bf Acknowledgments.} 
I thank Drs. C. Robert O'Dell,  C\'esar Brice\~no, and Doug Johnstone
for many helpful comments resulting from careful reading of the manuscript.
We acknowledge support by NASA grant NNA04CC11A to
the CU Center for Astrobiology and NSF grant AST0407356.
I thank David Thiel, who a decade ago spent years working with 
me on  the cometary clouds and overall properties of the Orion/Eridanus
superbubble and its relationships to stars and gas in Orion.


\begin{thebibliography}{}

\bibitem[]{}
Bally, J., Langer, W. D., Stark, A. A., \& Wilson, R. W. 1987,
 ApJ, 312, L45 
 
\bibitem[]{}
Bally, J., Langer, W. D., Wilson, R. W., Stark, A. A., \& Pound, M. W.
  1991a, in {\it Fragmentation of Molecular Clouds and Star Formation}, IAU
  Symp. 147, eds. E. Falgarone, F. Boulanger, \& G. Duvert, Kluwer, Dordrecht, p. 11

\bibitem[]{}
Bally, J., Langer, W. D., \& Liu, W. 1991b, ApJ, 383, 645

\bibitem[]{}
Blaauw, A. 1991,  in {\it The Physics of Star Formation and Early
Stellar Evolution} eds. C.J. Lada \& N. D. Kylafis, Kluwer,  p. 125

\bibitem[]{}
Boboltz, D.~A., Fey, 
A.~L., Puatua, W.~K., Zacharias, N., Claussen, M.~J., Johnston, K.~J., 
\& Gaume, R.~A.\ 2007, \aj, 133, 906 

\bibitem[Brice\~no et al.(2007)]{briceno07}
Brice\~no C., Hartmann, L., Hern\'andez J., Calvet, N., Vivas A.~K., et al. 2007b, \apj, 661, 1119

\bibitem[]{}
Brown, A. G. A., de Geus, E. J., \& de Zeeuw, P. T. 1994, A\&A, 289, 101

\bibitem[]{}
Brown, A. G. A., Hartmann, D., \& Burton, W. B. 1995, A\&A, 300, 903

\bibitem[]{}
Burrows, D. N., Singh, K. P., Nousek, J. A., Garmire, G. P., \&
  Good, J. 1993, ApJ, 406, 97

\bibitem[]{}
Cowie, L. L., Songaila, A., \& York, D. G. 1979, ApJ, 230, 469

\bibitem[]{}
Dame, T. M., Ungerechts, H., Cohen, R., DeGeuss, E., Grenier, I.,
  May, J., Murphy, D., Nyman, L. A., \& Thaddeus, P. 1987, ApJ, 322, 706

\bibitem[]{}
Dame, T. M., Ungerechts, H., Cohen, R., de Geuss, E., Grenier, I.,
  May, J., Murphy, D., Nyman, L. A., \& Thaddeus, P. 1987, ApJ, 322, 706
  
\bibitem[]{}
Dame, T. M., Hartmann, D., \& Thaddeus, P. 2001, ApJ, 547, 752

\bibitem[]{}
De Zeeuw, P. T., Hoogerwerf, R., De Bruijne, J. H. J., Brown, A. G. A., 
   \& Blaauw, A 1999, AJ, 117, 354
   
\bibitem[]{}
Diehl, R., Cervino, M., Hartmann, D. H., \& Kretschmer, K.
2004, New Astron. Rev., 48, 81
   
 \bibitem[]{}  
Elmegreen, B. G. 1997a, in NASA Estes Park Conference on
   {\it Origins of Galaxies, Stars, Planets, \& Life},
    eds. C. E. Woodward, H. A. Thronson, \& J. M. Shull,
    ASP Conference Series, 1998

\bibitem[]{}
Elmegreen, B. G. 1997b, ApJ, 486, 944 
   
\bibitem[]{}
Genzel, R., Reid, M. J., Moran, J. M., \& Downes, D. 1981,
     ApJ, 244, 884
     
\bibitem[]{}
Gies, D.~R. \& Bolton, C.~T.\ 1986, ApJS, 61, 419 

\bibitem[]{}
Gies, D. R. 1987, ApJS, 64, 545

\bibitem[]{}
Green, D. A. 1991, MNRAS, 253, 350
 
\bibitem[]{}
Green, D. A. \& Padman, R. 1993, MNRAS, 263, 535

\bibitem[]{}
Gualandris, A., 
Portegies Zwart, S., \& Eggleton, P.~P.\  2004, \mnras, 350, 615
     
\bibitem[]{}
Harper, G. H., Brown, A., \& Guinan, E. F.\  2008, AJ, 135, 1430

\bibitem[]{}
Heiles, C. 1976, ApJ, 208, L137

\bibitem[]{}
Heiles, C. 1998, ApJ, 498, 689

\bibitem[]{}
Hillenbrand, L. A. 1997, AJ, 113, 1733

\bibitem[]{}
Hirota, T., Bushimata, T., Choi, Y.K., Honma, M., Imai,
H. et al. 2007, PASJ, 59, 897

  
\bibitem[]{}
Hog,  E., Fabricius,  C., Makarov,  V.V., Urban,  S., Corbin,  T.,
    Wycoff , G., Bastian,  U., Schwekendiek,  P.,  \& Wicenec,  A.
   2000, A\&A,  355, L27 
 
\bibitem[]{}
Hoogerwerf, R., de Bruijne, J.~H.~J., \& de Zeeuw, P.~T.\ 2001, \aap, 365, 49

\bibitem[]{}
Johnstone, D. \& Bally, J. 1999, ApJ, 510, L49 

\bibitem[]{}
Kharchenko,  N.V., Scholz,  R.-D., Piskunov,  A.E., Roeser,  S.,  \& Schilbach,  E.
2007, Astron. Nachr., 328, 889 

\bibitem[]{}
Lesh, J. R. 1968, ApJS, 17 371

\bibitem[]{}
Lindblad, P. O. 1967, Bull. Astron. Soc. Netherlands. 19, 34

\bibitem[]{}
Lindblad, P. O. 1973, A\&A, 24, 309

\bibitem[]{}
MacLow M. M. \&  McCray, R. 1987, ApJ, 324, 776

\bibitem[]{}
Maddalena, R. J., Morris, M., Moscowitz, J., \& Thaddeus, P. 
1986 ApJ, 303, 375.

\bibitem[]{}
McCray, R. \& Kafatos, M.\ 1987, ApJ, 317, 190

\bibitem[]{}
Menten, K. M., Reid, M. J., Forbrich, J., \& Brunthaler, A. 2007. AA, 474, 515

\bibitem[]{}
Parenago, P. P. 1954, Trudy Gosudarstvennogo Astronomischeskogo Instituta, 25, 1

\bibitem[]{}
Perryman, M. A. C., Lindegren, L., Kovalevsky, J., Hoeg,
E., Bastian, U. et al. 1997, A\&A, 323, L49

\bibitem[]{}
Poppel, W. G. L., Marronetti, P., \& Benaglia, P.  1994, A\&A, 287, 601

\bibitem[]{}
Reynolds, R. J. \& Ogden, P. M. 1979, ApJ, 229, 942

\bibitem[]{}
Sandstrom, K. M., Peek, J. E. G., Bower, G. C., Bolatto, A. D., \& Plambeck, R. L.
2007, ApJ, 667, 1161

\bibitem[]{}
Shull,  J. M. \& Saken, J . M. 1995, ApJ, 444, 663

\bibitem[]{}
Sivan, J. P. 1974, A\&AS, 16, 163

\bibitem[]{}
Spitzer, L. 1978, {\it Physical Processes in the
Interstellar Medium}, Wiley-Interscience

\bibitem[]{}
Stark, A.A., Gammie, C.F., Wilson, R.W., Bally, J.,
Linke, R.A., Heiles, C., \& Hurwitz, M.  1992, ApJS, 79, 77

\bibitem[]{}
Taylor, D. K., Dickman, R. L., \& Scoville, N. Z. 1987, ApJ, 315, 104

\bibitem[Walter et al.(1997)]{walter97} 
Walter, F. M., Wolk, S. J., Freyberg, M., Schmitt, J. H. M. M. 1997, 
Memorie della Societa Astronomia Italiana, 68, 1081

\bibitem[]{}
Walker, M. F. 1969, ApJ, 155, 447

\bibitem[]{}
Warren, W. H., \& Hesser, J. E. 1977, ApJS, 34, 115

\end{thebibliography}
\end{document}